# CECGSR: Circular ECG Super-Resolution

Honggui LI[1*], Zhengyang ZHANG[2], Dingtai LI[3], Sinan CHEN[4], Nahid MD LOKMAN HOSSAIN[5], Hantao LU[6], Ruobing WANG[7], Xinfeng XU[8], Yinlu QIN[9], Yuting FENG[10], Maria TROCAN[11], Dimitri GALAYKO[12], Amara AMARA[13], Mohamad SAWAN[14,15]

[1,2,4,5,6,7,8,9,10]School of Information and Artificial Intelligence, Yangzhou University, Yangzhou 225127, China

[3]Shanghai Qigong Research Institute, Shanghai University of Traditional Chinese Medicine, Shanghai 201203, China

[11]LISITE Research Laboratory, Institut Supérieur d'Électronique de Paris, Paris 75006, France

[12]Laboratoire d'Informatique de Paris 6, Sorbonne University, Paris 75020, France

[13]International Campus (Hangzhou), Beijing University of Aeronautics and Astronautics, Hangzhou 310000, China

[14]School of Engineering, Westlake University, Hangzhou 310024, China

[15]Polystim Neurotech Laboratory, Polytechnique Montreal, Montreal H3T1J4, Canada

[1]hgli@yzu.edu.cn, [2]mz120250945@stu.yzu.edu.cn, [3]22025067@shutcm.edu.cn, [4]mz120240994@stu.yzu.edu.cn, [5]mh23083@stu.yzu.edu.cn, [6]241301211@stu.yzu.edu.cn, [7]241303321@stu.yzu.edu.cn, [8]241301123@stu.yzu.edu.cn, [9]241301115@stu.yzu.edu.cn, [10]241301206@stu.yzu.edu.cn, [11]maria.trocan@isep.fr, [12]dimitri.galayko@sorbonne-universite.fr, [13]amara@buaa.edu.cn, [14]sawan@westlake.edu.cn, [15]mohamad.sawan@polymtl.ca

[*]Corresponding Author

**Abstract**: **Background and Objective**: The electrocardiogram (ECG) plays a crucial role in the diagnosis and treatment of various cardiac diseases. ECG signals suffer from low-resolution (LR) due to the use of convenient acquisition devices, as well as internal and external noises and artifacts. Classical ECG super-resolution (ECGSR) methods adopt an open-loop architecture that converts LR ECG signals to super-resolution (SR) ones. According to the theory of automatic control, a closed-loop framework exhibits superior dynamic and static performance compared with its open-loop counterpart. **Methods**: This paper proposes a closed-loop approach, termed circular ECGSR (CECGSR), which models the degradation process from SR ECG signals to LR ones. The negative feedback mechanism of the closed-loop system is based on the differences between the LR ECG signals. A mathematical loop equation is constructed to characterize the closed-loop infrastructure. The Taylor series expansion is employed to demonstrate the near-zero steady-state error of the proposed method. A Plug-and-Play strategy is considered to establish the SR unit of the proposed architecture, leveraging any existing advanced open-loop ECGSR methods. This paper also presents Transformer model based open-loop ECGSR and closed-loop CECGSR algorithms. **Results**: Simulation experiments on both noiseless and noisy subsets of the Physikalisch-Technische Bundesanstalt-Extra Large (PTB-XL) datasets demonstrate that the proposed CECGSR outperforms state-of-the-art open-loop ECGSR algorithms in the reconstruction performance of ECG signals. **Conclusions**: The proposed method will efficiently enrich ECG signal details and remove ECG signal artifacts in clinical applications.

Series Expansion, Steady-State Error, Plug-and-Play

## 1 Introduction

Electrocardiography (ECG) employs multiple lead electrodes to capture cardiac electrical signals during the atrial and ventricular contraction and relaxation phases [1]. ECG can be utilized to diagnose and monitor a wide range of cardiovascular diseases, including cardiac arrhythmias, ventricular arrhythmias, atrial fibrillation, atrial flutter, myocarditis, myocardial infarction, myocardial fibrosis, myocardial ischemia, hypertrophic cardiomyopathy, pericarditis, cardiac amyloidosis, cardiac dysregulation, ventricular dysfunction, cardiac arrest, heart failure, sudden cardiac death, coronary artery disease, mitral valve disease, coronary vasospasm, tetralogy of Fallot, Brugada syndrome, Wolff-Parkinson-White syndrome, and so on.

ECG offers several advantages, such as non-invasiveness, cost-effectiveness, simplicity, convenience, versatility, accessibility, reliability, and functionality [2]. However, ECG has certain limitations, such as variability, uniqueness, low-resolution (LR), susceptibility to high levels of noises and artifacts, the need for prolonged recording, and limited diagnostic accuracy. Therefore, it is necessary to enhance the resolution of ECG signals for advanced applications, including ECG signal classification, recognition, interpretation, understanding, and so forth.

ECG super-resolution (ECGSR) reconstructs high-resolution (HR) or high-definition ECG signals from LR ECG signals [3]. ECGSR methods can be categorized into three types: model-based methods, learning-based methods, and hybrid methods. Model-based methods recover super-resolution (SR) ECG signals using predefined mathematical models, such as numerical interpolation, total variation, wavelet transform (WT), sparse representation, low-rank representation, and etc. Singhal S et al utilized Superlet transform, a HR technique, to reshape the 1-D ECG signal into a 2-D ECG signal [4]. Learning-based methods reconstruct SR ECG signals by leveraging modern deep learning-based stacked neural network architectures, such as autoencoder (AE), denoising autoencoder (DAE), variational autoencoder, convolution neural network (CNN), convolutional AE (CAE), denoising CAE (DCAE), generative neural network (GAN), transformer (TF) model, diffusion model, flow model, autoregressive model, Mamba model, and etc. Lomoio Ugo et al employed a DCAE for ECGSR [5]. Chen Tsai-Min et al proposed a cloud-based deep learning algorithm for ECGSR [6]. Jie Lin et al presented a Mamba model for ECGSR [7]. Hybrid methods refer to the integration of model-based and learning-based methods.

The generalization performance of learning-based methods is constrained by the inherent differences between training and testing samples [8]. The out-of-sample performance is consistently inferior to the in-sample performance. The performance degradation of out-of-sample data cannot be resolved solely through training. Classical ECGSR adopts an open-loop architecture that maps LR ECG signals to SR ECG signals. According to the theory of automatic control, closed-loop architectures outperform their open-loop counterparts in both dynamic and static performance [9]. Hence, this paper proposes a close-loop ECGSR method, termed Circular ECGSR (CECGSR), to enhance the generalization performance of pretrained open-loop learning-based ECGSR models. CECGSR is a hybrid method that combines learning-based method and the theory of automatic control. CECGSR is composed of a pretrained open-loop ECGSR model and a degradation model from SR samples to LR samples. Negative feedback mechanism of the closed-loop system relies on the differences between SR ECG signals. CECGSR is characterized and analyzed using a nonlinear loop equation. The steady-state error of CECGSR is approximately zero, which is mathematically proven by Taylor series expansion. CECGSR exhibits the property of Plug-and-Play (PnP), as it can take advantage of any existing advanced pretrained open-loop ECGSR models [10]. At the same time, brand-new TF model based open-loop ECGSR and closed-loop CECGSR infrastructures are also raised.

The proposed CECGSR is an extension of our previous work on generalized image restoration, including circular image SR, circular image compressed sensing (CS), and circular image compression [11-13]. The proposed CECGSR depends on the negative feedback principle based on the differences between degraded LR ECG signals, while our previous work depends on the negative feedback principle based on the differences between non-degraded HR images.

The key innovations of this paper are listed as follows:

- Closed-loop architecture of CECGSR incorporating pretrained and degraded models;
- TF model based open-loop ECGSR and closed-loop CECGSR frameworks;
- Negative feedback structure built on the differences between LR ECG signals;
- Nonlinear loop equation of CECGSR for mathematical description and analysis;
- Near-zero steady state error of CECGSR proven via Taylor series approximation;
- PnP property of CECGSR based on any existing pretrained open-loop ECGSR models;
- Remarkable enhancement in the generalization performance of CECGSR.

The remainder of this paper is organized as follows. Section 2 summarizes the related work, section 3 establishes the theoretical foundation, section 4 elaborates on the simulation experiments, and section 5 draws the corresponding conclusions.

## 2 Related-Work

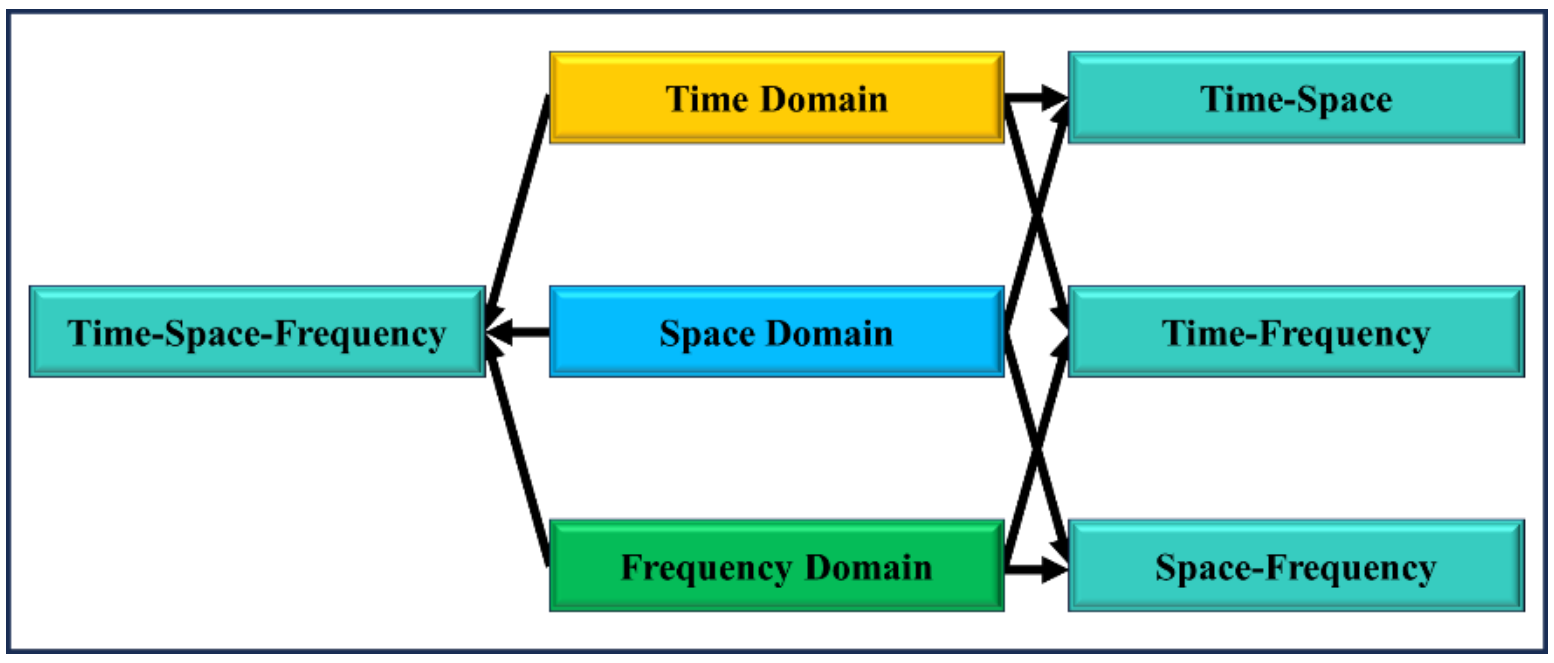

**Figure 1. Various domains of ECGSR.**

ECGSR can be implemented across multiple domains, including the time domain, space domain, frequency domain, time-space domain, time-frequency domain, space-frequency domain, time-space-frequency domain, and so on. The various domains of ECGSR are illustrated in Figure 1.

Time domain ECGSR can be directly gained by HR acquisition hardware, multiscale network, temporal network, digitization, data augmentation (DA), data synthesis, and etc. Luiz Luiz E. et al proposed a HR ECG acquisition system with 24-bit precision [14]. Lin Junye et al presented a multiscale CNN feature fusion bidirectional long-short-term-memory (LSTM) network for ECG based diagnosis [15]. Reddy C Kishor Kumar et al raised a multiscale convolutional LSTM-dense network for ECG signal classification [16]. Zhou Feiyan put forward a multiscale and hierarchical-feature CNN for ECG signal classification [17]. Huang Zhen-Zhen et al come up with a temporal convolutional GAN for ECG signal extraction [18]. Demolder Anthony et al proposed a method of high-precision ECG digitization using artificial intelligence (AI) [19]. Herman Robert et al presented AI-enabled ECG digitization and interpretation of occlusion myocardial infarction [20]. Li Yifan et al raised a DA approach via small-scale longitudinal compression and stretching to generate more authentic ECG samples [21]. Lim Jeonghwa et al put forward a DA method for synthesizing data while preserving accurate labels [22]. Adib Edmond et al come up with a synthetic ECG generation algorithm using generative neural networks [23]. Zanchi Beatrice et al summarized synthetic ECG signal generation methods [24].

Time domain ECGSR can be indirectly gained by filtering, denoising, enhancement, artifacts removal, and etc. Zhang Juya et al overviewed ECG noise interference suppression algorithms [25]. Dhas D. Edwin et al proposed an ECG artifacts removal method based on adjacent channel weight dependable recursive least square adaptive filter [26]. Yadav Nilesh Kumar et al presented a novel approach for ECG artifacts removal using multistage least-mean-square adaptive finite-impulse-response filter [27]. Li Shun et al raised an ECG noise-reduction method using complete-ensemble-empirical-mode-decomposition with adaptive-noise, Tuna-swarm-optimization, and stacked-sparse-autoencoder [28]. Qu Tianxiang et al put forward a hybrid motion artifacts removal method for ECG signal [29]. Alugonda Rajani et al come up with a CNN based ECG denoising with multivariate-intrinsic-empirical-mode-decomposition filtering [30]. Chakrabarty Bulty et al proposed optimized deep learning approach and smoothing filter for ECG denoising [31]. Yenila Firna et al enhanced ECG images using wave translation algorithm with continuous WT (CWT) [32]. Ullah Hadaate et al presented an ECG motion artifacts reduction method with 2D CAE [33]. Al-azzawi Nada et al raised low-frequency textile-based resistor-capacitor filters for ECG noise reduction [34]. Tiwari Vibha et al put forward an efficient sparse code shrinkage technique for ECG denoising [35]. Zhang Chenhua et al come up with an ECG denoising method based on variational mode decomposition and recursive least square [36]. Almohimeed Ibrahim et al proposed an ECG motion artifacts mitigation method using nonlinear autoregressive networks with cat swarm optimization and fractional calculus [37]. Shaheen Ahmed et al presented fully-gated DAE for ECG artifacts reduction [38]. Lai Shin-Chi et al raised compact shortcut DAE for ECG signal denoising [39]. Li Zhiyuan et al put forward a novel ECG denoising method using dual-path diffusion model [40]. Ge Liang et al come up with a joint ECG denoising method with improved wavelet thresholding and singular spectrum analysis [41]. Ramadhan Alif Wicaksana proposed an ECG denoising method based on CAE with sequential and channel attention [42]. Bentaleb Dounia et al presented multi-criteria Bayesian optimization of empirical mode decomposition and hybrid filters fusion for ECG denoising [43].

Time domain ECGSR can also be indirectly gained by ECG signal compression and CS. ECG signal compression reconstructs HR samples from related LR compressed ones. ECG signal CS recovers HR samples from down-sampled ones. Erna G. et al proposed an ECG compression method with lossless compression, run-length coding, and Golomb-Rice encoding [44]. Spagnolo Fanny et al presented a novel ECG CS -based cryptosystem [45]. Kumar Subramanyam Shashi raised a sparse ECG recovery approach for hybrid CS framework [46]. Huang Guoxing et al put forward an improved two-channel finite rate of innovation sampling framework and a modified annihilating filter reconstruction algorithm for ECG signal CS [47].

Space domain ECGSR can be attained through multi-channel acquisition hardware, multi-dimension ECG signal, ECG signal reconstruction, ECG-imaging, cardiac digital twins, and so on. Luiz Luiz E. et al proposed a HR ECG acquisition system with 8 channels [14]. Mahabadi Amir A. et al presented a novel 3D ECG algorithm for cardiac amyloidosis [48]. Obianom Ekenedirichukwu N. et al utilized as little as one lead and as much as four leads for reconstructing of the other leads [49]. Chen Hong et al rebuilt ECG signals from non-contact Ballistocardiogram signals [50]. Cui Renwei et al implemented non-contact reestablishment of ECG signals using frequency-modulated-continuous-wave radar and sequence-to-sequence-and-attention network [51]. Epmoghaddam Dorsa et al reconstructed 12-lead ECG signal from reduced lead sets using an encoder-decoder CNN [52]. Zhang Yuanyuan et al raised a robust radar-based ECG signal restoration based on multitask learning framework with eccentric gradient alignment [53]. El Ghebouli Ayoub et al automatically detected the location of the electrodes used for body surface potential maps leveraging 3D depth sensing camera [54]. Shenoy Nikhil et al put forward a novel 3D camera-based ECG-imaging system for electrode position discovery and heart-torso registration [55]. Li Lei et al come up with a personalized topology-informed localization of standard 12-lead ECG electrode placement from incomplete cardiac MRIs for efficient cardiac digital twins [56].

Frequency domain ECGSR can be obtained via high-frequency acquisition hardware and spectral enhancement. Luiz Luiz E. et al proposed a HR ECG acquisition system with 1-4 kHz sampling rate [14]. Muralidharan Pooja et al presented a spectral whitening method by removing the information related to the normal functioning for improving epileptic seizure prediction [57]. Vivar-Estudillo Guillermina et al raised spectral feature extraction method using the unbiased-finite-impulse-response iterative smoother algorithm for ECG signal classification [58].

Time-space domain ECGSR can potentially be achieved by integrating time-space information of ECG signals. Zou Jiawen et al proposed a novel spatiotemporal fusion deep CS method [59]. A channel attention mechanism is introduced to prioritize key features across different ECG leads and a multi-head mixed-attention mechanism is introduced to capture long-term temporal dependencies [59]. Abousaber Inam et al adopted spatiotemporal TFs and hybrid loss optimization for explainable ECG classification [60]. Liu Chien-Liang et al employed spatiotemporal attention residual networks for multi-label ECG abnormality classification [61].

Time-frequency domain ECGSR can efficiently be achieved by HR time-frequency representation and the combination of time-frequency features. Li Rui et al proposed a time synchro-squeezing generalized W transform to characterize the features of ECG signals with HR in both time and frequency domains [62]. Malleswari Pinjala N. et al wielded CWT to construct 2D ECG spectrograms, time-frequency visualization of ECG signals, for arrhythmia classification [63]. Khatar Zakaria et al presented a gradient-weighted-class-activation-mapping guided dynamic weighted fusion of temporal and frequency features of ECG signals [64]. ECG signals are firstly transformed into 2D Gramian angular field matrices, then temporal dynamics are captured by Gramian angular summation fields, and frequency dependencies are extracted by CWT [64].

Time-space-frequency domain ECGSR can possibly be achieved by merging time-frequency-space information of ECG signals. Liu Chien-Liang proposed a multimodal fusion model that integrates time, frequency, and space domain features of ECG signals [65]. A hierarchical dual-attention vision TF was presented for extracting detailed spatial-temporal features and a deep residual network was presented for capturing frequency domain representations via wavelet packet coefficients [65].

In summary, this paper proposes hybrid time domain CECGSR by fusing pretrained learning-based ECGSR and automatic control theory.

## 3 Theory

### 3.1 Mathematical Notations

The mathematical notations employed in this paper are presented in Table 1 below.

**Table 1. Mathematical notations.**

| Notation | Meaning |
|---|---|
| LR / SR / PP | low-resolution / super-resolution / pre-position functions |
| RG / DS | regularization / down-sampling functions |
| SL / LS | nonlinear functions |
| PSNR / SSIM | peak signal-to-noise ratio / structural similarity functions |
| $\mathbf{P}$ / $\mathbf{Q}$ / $\mathbf{H}$ | matrices of Taylor series coefficients / 2-order and high-order terms |
| $\mathbf{I}$ / $\mathbf{\Lambda}$ | identity matrix / matrix of multiplication coefficients |
| $\mathbf{s}_{l0}$ / $\mathbf{s}_{l}$ | original / generated LR ECG signals |
| $\mathbf{s}_{0}$ / $\mathbf{s}_{s}$ | original SR ECG signal / reconstructed SR ECG signal |
| $\mathbf{s}_{e}$ / $\mathbf{s}_{p}$ / $\mathbf{s}_{m}$ | LR error signal / LR PP signal / LR output signal of multiplier |
| $\xi$ / $\mathbf{e}$ | SR signal / steady-state error |
| $\mathbf{n}$ / $\mathbf{0}$ / $\boldsymbol{\sigma}$ | signal / mean / standard deviation of random noise |
| $\mu$ / $\sigma$ | element-wised mean / standard deviation or covariance |
| C / r / f | total number of channels / subsampling rate / SR factor |
| d / D | dimension of multi-channel LR / SR sample |
| $d_c$ / $D_c$ | dimension of single-channel LR / SR sample |
| $\eta$ / $\lambda$ | constant coefficients |
| N / T | order of partial derivative / total number of iterations |
| t / $\Delta$t | time / time interval |

### 3.2 Recall of Classical ECGSR

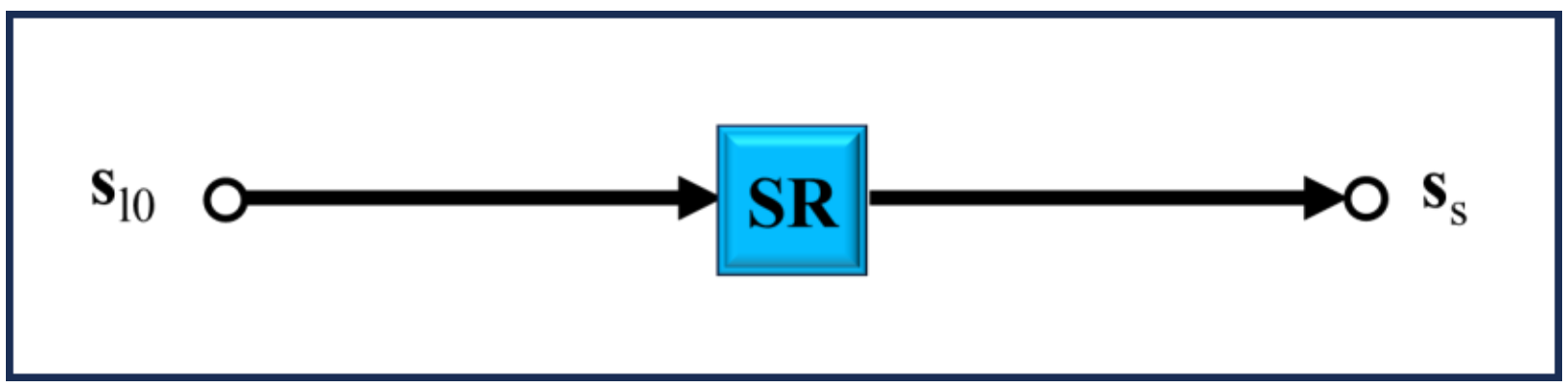

**Figure 2. The architecture of the classical ECGSR.**

The architecture of the classical ECGSR is illustrated in Figure 2. $\mathbf{s}_{l0}$ is the original LR ECG signal and $\mathbf{s}_{s}$ is the SR ECG signal. The SR module reconstructs $\mathbf{s}_{s}$ from $\mathbf{s}_{l0}$. The classical ECGSR can be formulated as the following mathematical equations:

$$\mathbf{s}_s = \mathrm{SR}(\mathbf{s}_{l0})$$
$$d = Cd_c, D = CD_c, r = \frac{d}{D} = \frac{d_c}{D_c}, f = \frac{1}{r} = \frac{D}{d} = \frac{D_c}{d_c}, \text{s.t. } r, f \in R; \mathbf{s}_{l0} \in R^d; \mathbf{s}_s \in R^D, \quad (1)$$

where: d is the dimension of multi-channel LR sample, C is the total number of channels, $d_c$ is the dimension of single-channel LR sample, D is the dimension of multi-channel SR sample, $D_c$ is the dimension of single-channel SR sample, r is the subsampling rate, and f is the SR factor.

The classical ECGSR can be solved via the following minimization problem:

$$\mathbf{s}_s^* = \arg\min_{\mathbf{s}_s} \frac{1}{2}\left\| \mathbf{s}_{l0} - \mathrm{LR}(\mathbf{s}_s) \right\|_2^2 + \eta \cdot \mathrm{RG}(\mathbf{s}_s)$$
$$\text{s.t. } \mathrm{LR}(\mathbf{s}_s) = \mathrm{DS}(\mathbf{s}_s) + \mathbf{n}, \mathbf{n} \in N(\mathbf{0}, \boldsymbol{\sigma}); \mathbf{0}, \boldsymbol{\sigma} \in R^d; \eta \in R, \quad (2)$$

where: LR function degrades $\mathbf{s}_s$ to $\mathbf{s}_{l0}$; LR function can be simple down-sampling with noise for classical SR and learnable degradation process for blind or real-world SR; η denotes a constant; RG function denotes the regularization of $\mathbf{s}_s$, such as total variation, sparsity representation, low-rank representation, and deep learning-based priors; DS function down-samples $\mathbf{s}_s$, such as down-sampling interpolation; **n** denotes the random noise with norm distribution; **0** denotes the mean of **n**; **σ** denotes the standard deviation of **n**.

### 3.3 Proposed Architecture of CECGSR

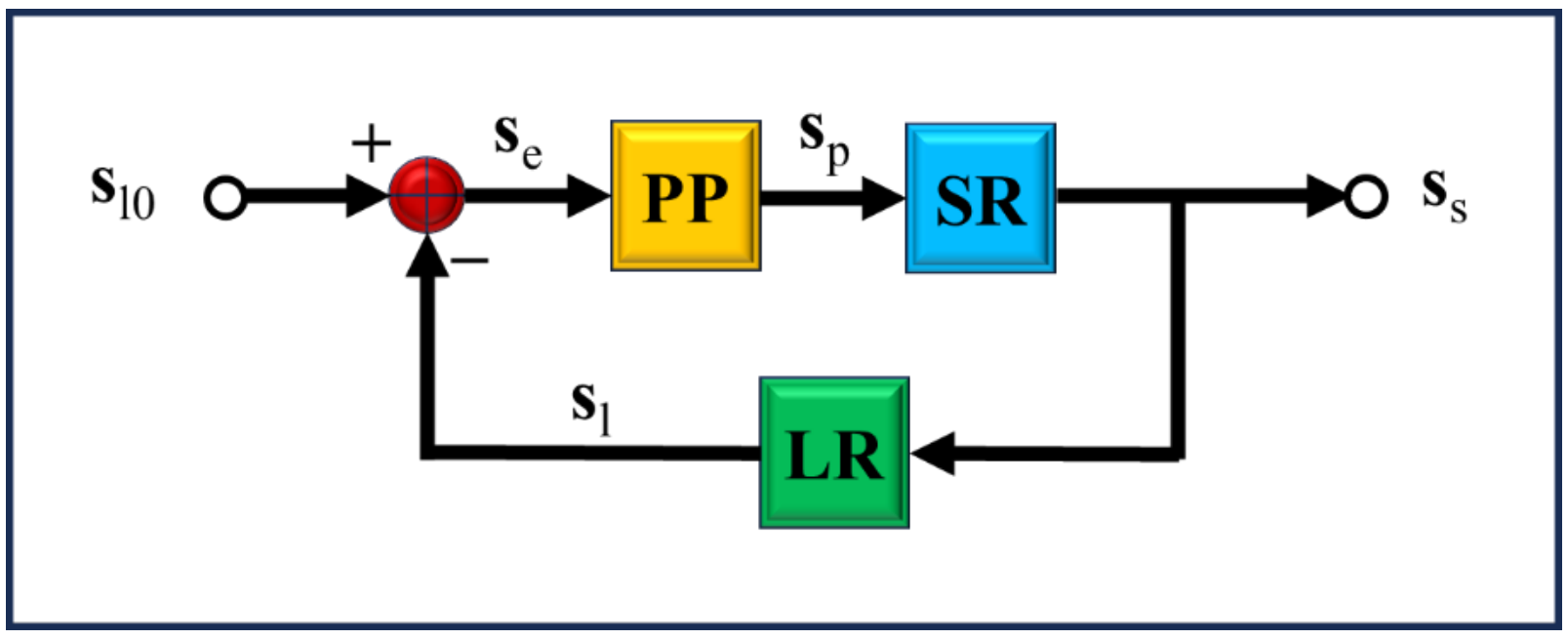

**Figure 3. The proposed architecture of CECGSR.**

The proposed architecture of CECGSR is depicted in Figure 3. $\mathbf{s}_l$ is the generated LR ECG signal, $\mathbf{s}_e$ is the LR error signal, and $\mathbf{s}_p$ is the LR pre-position (PP) signal. $\mathbf{s}_{l0}$ is the architecture input and $\mathbf{s}_s$ is the architecture output. The adder module introduces negative feedback, compares $\mathbf{s}_{l0}$ with $\mathbf{s}_l$, and produces $\mathbf{s}_s$. The negative feedback mechanism depends on the differences between LR ECG signals. The PP module transforms $\mathbf{s}_e$ into $\mathbf{s}_p$. The PP module can be either a traditional proportional-integral (PI) unit or a modern fuzzy unit. The SR module reconstructs $\mathbf{s}_s$ from $\mathbf{s}_p$. The SR module can adopt any existing or proposed pretrained open-loop SR algorithms. The LR module degrades $\mathbf{s}_s$ to $\mathbf{s}_l$. The LR module is known to perform down-sampling with noise in classical SR and is learnable in blind or real-world SR scenarios. Nearest-neighbor interpolation, bilinear interpolation, bicubic interpolation, and spline interpolation can be employed to simulate the down-sampling process of the LR module in classical SR.

According to the automatic control theory, when the closed-loop system is steady, $\mathbf{s}_e$ is close to zero, $\mathbf{s}_l$ is close to $\mathbf{s}_{l0}$, $\mathbf{s}_s$ is close to the original SR ECG signal $\mathbf{s}_0$, and perfect SR ECG signal is obtained.

### 3.4 Loop Equation of CECGSR

The loop equation of CECGSR can be described by the following mathematical equations:

$$\begin{cases} \mathbf{s}_e = \mathbf{s}_{l0} - \mathbf{s}_l \\ \mathbf{s}_p = \mathrm{PP}\left(\mathbf{s}_e\right) \\ \mathbf{s}_s = \mathrm{SR}\left(\mathbf{s}_p\right) \\ \mathbf{s}_l = \mathrm{LR}\left(\mathbf{s}_s\right) \end{cases} ,(3)$$

$$\Rightarrow \mathbf{s}_s = \mathrm{SR}\left(\mathrm{PP}\left(\mathbf{s}_{l0} - \mathrm{LR}\left(\mathbf{s}_s\right)\right)\right), \text{s.t. } \mathbf{s}_{l0}, \mathbf{s}_l, \mathbf{s}_e, \mathbf{s}_p \in \mathrm{R}^d; \mathbf{s}_s \in \mathrm{R}^D$$

The loop equation represents a nonlinear implicit equation about $\mathbf{s}_s$.

### 3.5 Stability of CECGSR

The stability of CECGSR refers to the ability of the closed-loop system to operate normally with a steady-state error approaching zero.

The aforementioned loop equation can be rephrased as follows:

$$\mathbf{s}_s = \mathrm{SL}\left(\mathbf{s}_s\right) = \mathrm{SR}\left(\mathrm{PP}\left(\mathbf{s}_{l0} - \mathrm{LR}\left(\mathbf{s}_s\right)\right)\right) ,(4)$$

where $\mathrm{SL}(\mathbf{s}_s)$ is a nonlinear function.

Assuming that $\mathrm{SL}(\mathbf{s}_s)$ has N+1 order partial derivatives at $\mathbf{s}_0$, its Taylor series expansion at $\mathbf{s}_0$ is:

$$\mathrm{SL}\left(\mathbf{s}_s\right) = \sum_{n=0}^{1} \mathbf{P}_n \left(\mathbf{s}_s - \mathbf{s}_0\right)^n + \mathbf{H}\left(\mathbf{s}_s\right), \mathbf{P}_n = \begin{bmatrix} p_{n;11} & p_{n;12} & \cdots & p_{n;1D} \\ p_{n;21} & p_{n;22} & \cdots & p_{n;2D} \\ \vdots & \vdots & \ddots & \vdots \\ p_{n;D1} & p_{n;D2} & \cdots & p_{n;DD} \end{bmatrix} \in \mathrm{R}^{D\times D}$$

$$p_{n;ij} = \begin{cases} p_{0;ij} = \dfrac{1}{D}\dfrac{1}{0!}\dfrac{\partial^{(0)}\mathrm{SL}_i\left(\mathbf{s}_s\right)}{\partial s_{s;j}^0}\bigg|_{\mathbf{s}_s=\mathbf{s}_0} = \dfrac{1}{D}\mathrm{SL}_i\left(\mathbf{s}_0\right), n=0 \\ p_{n;ij} = \dfrac{1}{n!}\dfrac{\partial^{(n)}\mathrm{SL}_i\left(\mathbf{s}_s\right)}{\partial s_{s;j}^n}\bigg|_{\mathbf{s}_s=\mathbf{s}_0}, n=1 \end{cases} ,(5)$$

$$\left(\mathbf{s}_s - \mathbf{s}_0\right)^n = \left[\left(s_{s;1} - s_{0;1}\right)^n \quad \left(s_{s;2} - s_{0;2}\right)^n \quad \cdots \quad \left(s_{s;D} - s_{0;D}\right)^n\right]^T$$

where: $\mathbf{P}_n$ denotes the matrix of Taylor series coefficients, $\mathbf{H}(s_s)$ is the 2-order and high-order terms, $SL_i(\mathbf{s}_s)$ denotes the i-th element of $SL(\mathbf{s}_s)$, $s_{s;j}$ denotes the j-th element of $\mathbf{s}_s$, ξ denotes a vector between $\mathbf{s}_0$ and $\mathbf{s}_s$, and $\xi_j$ denotes the j-th element of **ξ**.

Discarding $\mathbf{H}(s_s)$ of $SL(s_s)$, its linear approximation is:

$$\begin{gathered} SL(\mathbf{s}_s) \approx SL(\mathbf{s}_0) + \mathbf{P}_1(\mathbf{s}_s - \mathbf{s}_0) = \mathbf{P}_1(\mathbf{s}_s - \mathbf{s}_0) \\ \text{s.t. } SL(\mathbf{s}_0) = SR\left(PP\left(\mathbf{s}_{10} - LR(\mathbf{s}_0)\right)\right) = SR\left(PP\left(\mathbf{s}_{10} - \mathbf{s}_{10}\right)\right) = SR\left(PP(0)\right) = 0; \\ LR(\mathbf{s}_0) = \mathbf{s}_{10} \end{gathered} .(6)$$

The local linear approximation of loop equation at $\mathbf{s}_0$ is:

$$\mathbf{s}_s = SL(\mathbf{s}_s) \approx \mathbf{P}_1(\mathbf{s}_s - \mathbf{s}_0) .(7)$$

SR ECG signal $\mathbf{s}_s$ can be resolved as follows:

$$\mathbf{s}_s = (\mathbf{P}_1 - \mathbf{I})^{-1} \mathbf{P}_1 \mathbf{s}_0 ,(8)$$

where **I** represents the identity matrix.

The necessary and sufficient condition for errorless reconstruction is as follows:

$$(\mathbf{P}_1 - \mathbf{I})^{-1} \mathbf{P}_1 = \mathbf{I} \Leftrightarrow \mathbf{s}_s = \mathbf{s}_0 .(9)$$

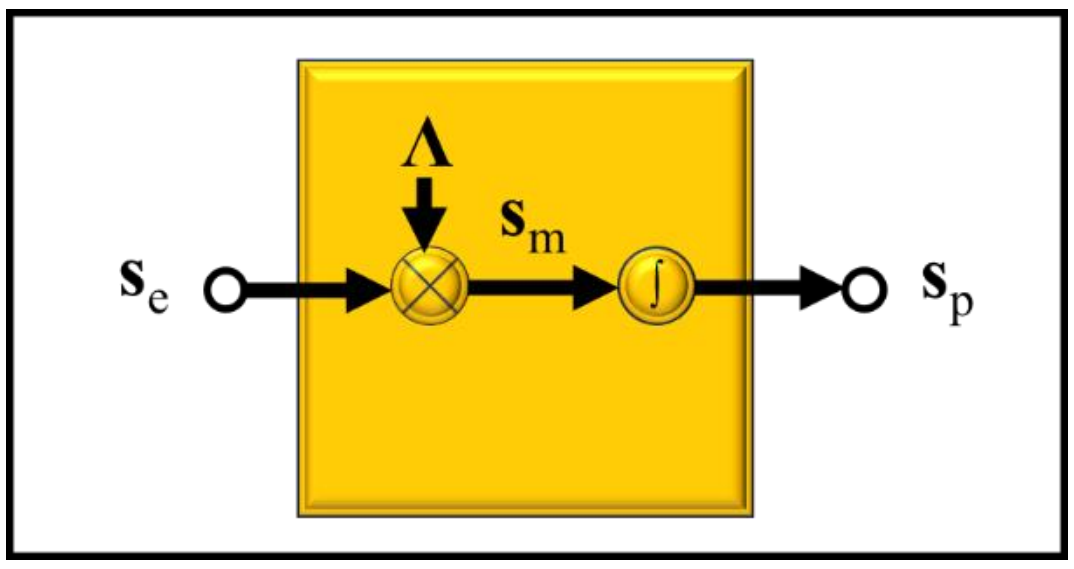

**Figure 4. Traditional PI control unit.**

The aforementioned condition is ideal and cannot be achieved in practice. To derive a practical condition, it is necessary to simplify the PP module. It is reasonable to assume that the PP module adopts traditional PI control structure in Figure 4 where **Λ** is the matrix of multiplication coefficients and $\mathbf{s}_m$ is the output of multiplier.

The traditional PI control unit can be depicted by the following mathematical expressions:

$$\mathbf{s}_m(t) = \mathbf{\Lambda}(t)\mathbf{s}_e(t), \mathbf{s}_p(t) = \int_0^t \mathbf{s}_m(t)dt + \mathbf{s}_p(0), \mathbf{\Lambda}(t) \in R^{d\times d} ,(10)$$

where: t is the time and $\mathbf{s}_p(0)$ is the initial value of integral.

According to Figure 4, the loop equation can be rewritten as follows:

$$\mathbf{s}_l(t) = \mathrm{LR}(\mathbf{s}_s(t)) = \mathrm{LR}(\mathrm{SR}(\mathbf{s}_p(t))) = \mathrm{LS}(\mathbf{s}_p(t)), \quad (11)$$

where $\mathrm{LS}(\mathbf{s}_p(t))$ denotes a nonlinear function.

Supposing that $\mathrm{LS}(\mathbf{s}_p(t))$ has (N+1)-order partial derivatives at $\mathbf{s}_p(0)$, its Taylor series expansion at $\mathbf{s}_p(0)$ is:

$$\mathrm{LS}(\mathbf{s}_p(t)) = \sum_{n=0}^{1} \mathbf{Q}_n (\mathbf{s}_p(t) - \mathbf{s}_p(0))^n + \mathbf{H}(\mathbf{s}_p(t))$$

$$\mathbf{Q}_n = \begin{bmatrix} q_{n;11} & q_{n;12} & \cdots & q_{n;1d} \\ q_{n;21} & q_{n;22} & \cdots & q_{n;2d} \\ \vdots & \vdots & \ddots & \vdots \\ q_{n;d1} & q_{n;d2} & \cdots & q_{n;dd} \end{bmatrix} \in R^{d\times d}$$

$$q_{n;ij} = \begin{cases} q_{0;ij} = \dfrac{1}{d}\dfrac{1}{0!}\left.\dfrac{\partial^{(0)}\mathrm{LS}_i(\mathbf{s}_p(t))}{\partial s_{p;j}^0(t)}\right|_{\mathbf{x}_p(t)=\mathbf{x}_p(0)} = \dfrac{1}{d}\mathrm{LS}_i(\mathbf{s}_p(0)), n=0 \\ q_{n;ij} = \dfrac{1}{n!}\left.\dfrac{\partial^{(n)}\mathrm{LS}_i(\mathbf{s}_p(t))}{\partial s_{p;j}^n(t)}\right|_{\mathbf{s}_p(t)=\mathbf{s}_p(0)}, n=1 \end{cases}, \quad (12)$$

$$(\mathbf{s}_p(t) - \mathbf{s}_p(0))^n = \left[ (s_{p;1}(t) - s_{p;1}(0))^n \quad (s_{p;2}(t) - s_{p;2}(0))^n \quad \cdots \quad (s_{p;d}(t) - s_{p;d}(0))^n \right]^T$$

where: $\mathbf{Q}_n$ represents the matrix of Taylor series coefficients, $\mathbf{H}(s_p(t))$ is the 2-order and high-order terms, $\mathrm{LS}_i(\mathbf{s}_p(t))$ represents the i-th element of LS ($\mathbf{s}_p(t)$), and $s_{p;j}(t)$ represents the j-th element of $\mathbf{s}_p(t)$.

Discarding $\mathbf{H}(s_p(t))$ of $\mathrm{LS}(\mathbf{s}_p(t))$, its linear approximation is:

$$\mathrm{LS}(\mathbf{s}_p(t)) \approx \mathrm{LS}(\mathbf{s}_p(0)) + \mathbf{Q}_1(\mathbf{s}_p(t) - \mathbf{s}_p(0)). \quad (13)$$

The loop equation can be linearly approximated as follows:

$$\mathbf{s}_l(t) \approx \mathrm{LS}(\mathbf{s}_p(0)) + \mathbf{Q}_1(\mathbf{s}_p(t) - \mathbf{s}_p(0)). \quad (14)$$

The loop equation can further be rewritten as follows:

$$\mathbf{s}_1(t) \approx \mathrm{LS}(\mathbf{s}_p(0)) + \mathbf{Q}_1\left(\int_0^t \mathbf{\Lambda}(t)\mathbf{s}_e(t)dt\right).(15)$$

Assuming that time interval $\Delta t$ is greater than zero, $\mathbf{s}_1(t+\Delta t)$ can be gained as follows:

$$\mathbf{s}_1(t+\Delta t) = \mathbf{s}_1(t) + \mathbf{Q}_1\left(\int_t^{t+\Delta t} \mathbf{\Lambda}(t)\mathbf{s}_e(t)dt\right).(16)$$

Subtracting both sides of aforementioned equation from $\mathbf{s}_{10}$, the following equation can be obtained:

$$\mathbf{s}_{10} - \mathbf{s}_1(t+\Delta t) = \mathbf{s}_{10} - \mathbf{s}_1(t) - \mathbf{Q}_1\left(\int_t^{t+\Delta t} \mathbf{\Lambda}(t)\mathbf{s}_e(t)dt\right).(17)$$

According to the definition of error signal $\mathbf{s}_e$, the following equation can be attained:

$$\mathbf{s}_e(t+\Delta t) = \mathbf{s}_e(t) - \mathbf{Q}_1\left(\int_t^{t+\Delta t} \mathbf{\Lambda}(t)\mathbf{s}_e(t)dt\right).(18)$$

Supposing that $\Delta t$ is close to zero, the following equation can be achieved:

$$\mathbf{s}_e(t+\Delta t) \approx \mathbf{s}_e(t) - \mathbf{Q}_1\mathbf{\Lambda}(t)\mathbf{s}_e(t)\Delta t = (\mathbf{I} - \mathbf{Q}_1\mathbf{\Lambda}(t)\Delta t)\mathbf{s}_e(t), \text{s.t. } \Delta t \to 0.(19)$$

By taking the norm of both sides of the aforementioned equation, the following inequality can be derived under certain conditions:

$$\left\|\mathbf{s}_e(t+\Delta t)\right\|_2 \le \left\|\mathbf{I} - \mathbf{Q}_1\mathbf{\Lambda}(t)\Delta t\right\|_F \left\|\mathbf{s}_e(t)\right\|_2, \text{s.t. } \Delta t \to 0,(20)$$

where: subscript 2 is 2-norm and subscript F is Frobenius-norm.

The condition that satisfies the aforementioned inequality is the following one:

$$\left\|\mathbf{I} - \mathbf{Q}_1\mathbf{\Lambda}(t)\Delta t\right\|_F < 1.(21)$$

Suitable $\mathbf{Q}_1$ and $\mathbf{\Lambda}$ can always be chosen to meet the aforementioned condition. If $\mathbf{Q}_1$ and $\mathbf{\Lambda}$ are proportional to identity matrix, the following equations can be obtained:

$$\begin{cases} \mathbf{Q}_1 = q_1\mathbf{I} \\ \mathbf{\Lambda}(t) = \lambda\mathbf{I} \end{cases}, \text{s.t. } q_1 = \frac{1}{d^2}\sum_{i=1}^{d}\sum_{j=1}^{d} q_{1;ij},(22)$$

where: $q_1$ denotes the average of the elements of $\mathbf{Q}_1$, and λ denotes a coefficient.

λ can be selected according to following inequality:

$$\begin{aligned} &\left\|\mathbf{I}-\mathbf{Q}_1\boldsymbol{\Lambda}(t)\Delta t\right\|_F=\left\|\mathbf{I}-q_1\mathbf{I}\cdot\lambda\mathbf{I}\cdot\Delta t\right\|_F=\left\|(1-\Delta t\cdot q_1\cdot\lambda)\mathbf{I}\right\|_F \\ &=\left|1-\Delta t\cdot q_1\cdot\lambda\right|\cdot\left\|\mathbf{I}\right\|_F=\left|1-\Delta t\cdot q_1\cdot\lambda\right|\cdot\sqrt{d}<1 \\ &\Rightarrow\begin{cases}\dfrac{1-\frac{1}{\sqrt{d}}}{\Delta t\cdot q_1}<\lambda<\dfrac{1+\frac{1}{\sqrt{d}}}{\Delta t\cdot q_1}, q_1>0 \\ \dfrac{1+\frac{1}{\sqrt{d}}}{\Delta t\cdot q_1}<\lambda<\dfrac{1-\frac{1}{\sqrt{d}}}{\Delta t\cdot q_1}, q_1<0\end{cases}\end{aligned} \quad .(23)$$

If the aforementioned condition is satisfied, the following inequality can be derived:

$$\left\|\mathbf{s}_e(t+\Delta t)\right\|_2<\left\|\mathbf{s}_e(t)\right\|_2 \quad .(24)$$

t is close to infinite in steady-state, then error signal $\mathbf{s}_e$ is approximate to zero:

$$\lim_{t\to\infty}\left\|\mathbf{s}_e(t)\right\|_2\to 0\Rightarrow\lim_{t\to\infty}\mathbf{s}_e(t)\to 0 \quad .(25)$$

Hence, $\mathbf{s}_l(t)$ is close to $\mathbf{s}_{l0}$:

$$\lim_{t\to\infty}\mathbf{s}_l(t)=\lim_{t\to\infty}\left(\mathbf{s}_{l0}-\mathbf{s}_e(t)\right)=\mathbf{s}_{l0}-\lim_{t\to\infty}\mathbf{s}_e(t)\to\mathbf{s}_{l0} \quad .(26)$$

Furthermore, $\mathbf{s}_s(t)$ is approximate to $\mathbf{s}_0$:

$$\left.\begin{aligned}\lim_{t\to\infty}\mathbf{s}_l(t)=\lim_{t\to\infty}\mathrm{LR}\left(\mathbf{s}_s(t)\right)=\mathrm{LR}\left(\lim_{t\to\infty}\mathbf{s}_s(t)\right) \\ \mathbf{s}_{l0}=\mathrm{LR}\left(\mathbf{s}_0\right)\end{aligned}\right\}\Rightarrow\lim_{t\to\infty}\mathbf{s}_s(t)\to\mathbf{s}_0 \quad .(27)$$

Finally, steady-state error $\mathbf{e}(t)$ is close to zero:

$$\begin{gathered}\mathbf{e}(t)=\mathbf{s}_0-\mathbf{s}_s(t) \\ \lim_{t\to\infty}\mathbf{e}(t)=\lim_{t\to\infty}\left(\mathbf{s}_0-\mathbf{s}_s(t)\right)=\mathbf{s}_0-\lim_{t\to\infty}\mathbf{s}_s(t)\to 0\end{gathered} \quad .(28)$$

At this point, the stability of CECGSR is proven, and the steady-state error approaches zero.

### 3.6 Negative Feedback

The negative feedback structure can be implemented on the differences between the LR or SR samples. The proposed CECGSR framework in Figure 3 depends on the differences between the LR samples. The CECGSR framework that relies on the differences between the SR samples is demonstrated in Figure 5, where $\mathbf{s}_{s0}$ is the initial reconstruction sample and $\mathbf{s}_p$ is the output SR sample. The proposed CECGSR architecture in Figure 5 is similar to our previous work on generalized image restoration [11-13]. The proposed architectures in Figures 3 and 5 are respectively named as CECGSR architecture I and II.

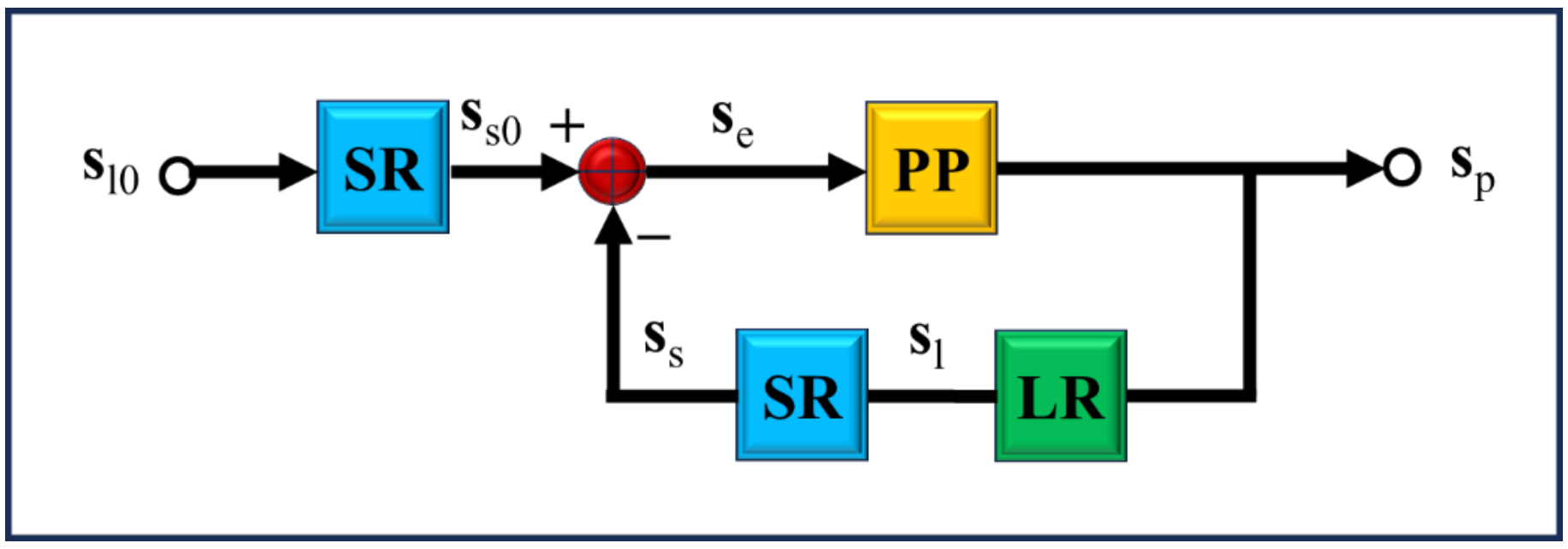

**Figure 5. CECGSR architecture based on the differences between the SR samples.**

### 3.7 Algorithm Description

The proposed CECGSR algorithm is illustrated in Figure 6, where T denotes the total number of iterations.

**Algorithm: CECGSR**
**Input:** $\mathbf{s}_{l0}$
**Initialization:**
  $t = 1$, $T$ = constant, $\lambda$ = constant, $\mathbf{s}_p(0)$ = constant
**while** $t <= T$
  compute $\Lambda(t)$ and $Q_1$ according to equation (22)
  compute $\mathbf{s}_l(t)$ according to equation (14)
  update $\mathbf{s}_s(t)$ according to equation (3)
  $t = t+1$
**end while**
**Output:** $\mathbf{s}_s(t)$

**Figure 6. Proposed algorithm of CECGSR.**

## 4 Experiments

### 4.1 Experimental Datasets

The details of the Physikalisch-Technische Bundesanstalt-Extra Large (PTB-XL) dataset are enumerated in Table 2 [5, 66]. The PTB-XL dataset contains five subsets: Conduction Disturbance (CD), Hypertrophy (HYP), Myocardial Infraction (MI), Normal (NORM), and ST/TC Change (STTC). The PTB-XL dataset includes two versions: a noiseless version and a noisy version. The noiseless version corresponds to the original PTB-XL dataset, while the noisy version is the modified PTB-XL dataset incorporating random noises and artifacts: electromyography (EMG), electrodermal activity (EDA), and baseline wander (BW).

**Table 2. Information of PTB-XL dataset.**

| Item | Subitem | Value |
|---|---|---|
| Total number of patients | N/A | 18869 |
| Sample point precision | N/A | 16 bits, 1 uV/LSB |
| Total number of leads | Limb leads (I, II, III, aVR, aVL, aVF) | 6 |
| | Periodical leads (V1~V6) | 6 |
| | Total number | 12 |
| Sample period | N/A | 5 s |
| Sampling rate | LR / SR | 50 / 500 Hz |
| Subsampling rate | N/A | 0.1 |
| SR factor | N/A | 10 |
| Sample length | LR / SR | 250 / 2500 points |
| Total number of samples | CD | 4650 |
| | HYP | 2616 |
| | MI | 10848 |
| | NORM | 19028 |
| | STTC | 5634 |
| | Total number | 42766 |
| Sample division | Percentage of training / testing samples | 90% / 10% |
| | Total percentage | 100% |
| Noise and artifacts | EMG / EDA / BW | N/A |
| Web links | https://github.com/UgoLomoio/DCAE-SR | N/A |
| | https://physionet.org/content/ptb-xl/1.0.3/ | N/A |

### 4.2 Competing Algorithms

**Table 3. Competing algorithms.**

| Algorithm | Web Link | Reference |
|---|---|---|
| CAESR | https://github.com/noahsafar/ECG-Autoencoder | [3] |
| CCAESR | N/A | proposed |
| DCAESR | https://github.com/UgoLomoio/DCAE-SR | [5] |
| CDCAESR | N/A | proposed |
| TFSR | N/A | proposed |
| CTFSR | N/A | proposed |

Three open-loop competing algorithms — classical convolutional autoencoder SR (CAESR) [3], classical denoising CAESR (DCAESR) [5], and proposed Transformer SR (TFSR) — along with three proposed closed-loop algorithms related thereto, namely circular CAESR (CCAESR), circular DCAESR (CDCAESR), and circular TFSR (CTFSR), are implemented to compare their reconstruction performance. The aforementioned algorithms are listed in Table 3 below. The proposed TFSR framework is shown in Figure 7 which contains four layers: local feature extraction, patching embedding, un-patching projection, and up-sampling head. Conv1d means 1D convolution and GELU means Gaussian error linear unit.

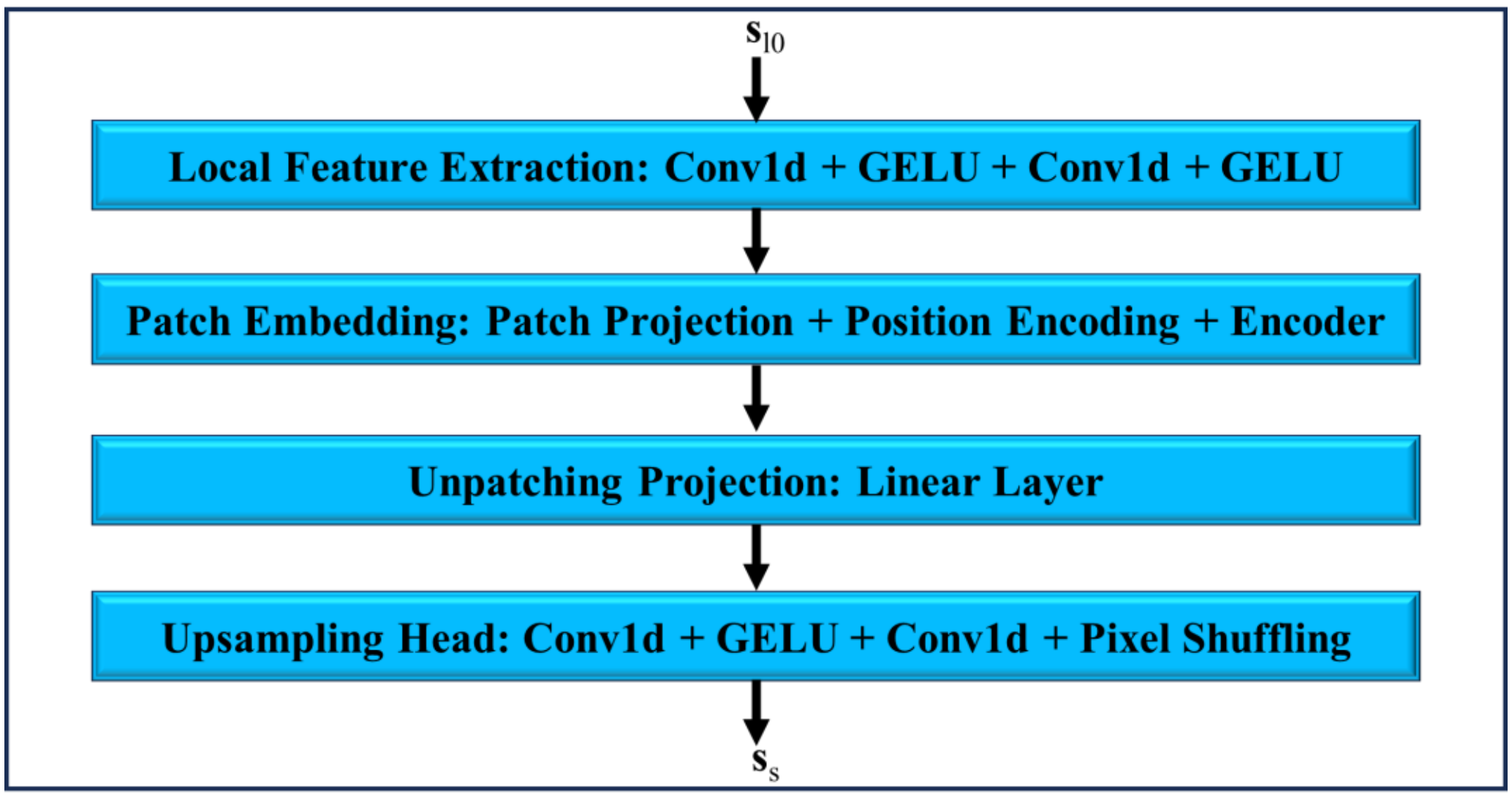

**Figure 7. Proposed framework of TFSR.**

### 4.3 Performance Metrics

Two performance metrics — peak signal-to-noise ratio (PSNR) and structural similarity (SSIM) — are adopted to evaluate the signal reconstruction capability of ECGSR and CECGSR. For the restoration SR signal $\mathbf{s}_s$ and the original SR signal $\mathbf{s}_0$, the definitions of PSNR and SSIM are depicted by the following mathematical formulas:

$$\mathrm{PSNR}(\mathbf{s}_s,\mathbf{s}_0)=10\lg\left(\frac{\mathrm{MAX}^2}{\frac{1}{D}\sum_{i=1}^{D}(s_{si}-s_{0i})^2}\right)(\mathrm{dB}), \quad (29)$$

$$\mathrm{SSIM}(\mathbf{s}_s,\mathbf{s}_0)=\frac{\left(2\mu_{s_s}\mu_{s_0}+(0.01\times\mathrm{MAX})^2\right)\left(2\sigma_{s_s s_0}+(0.03\times\mathrm{MAX})^2\right)}{\left(\mu_{s_s}^2+\mu_{s_0}^2+(0.01\times\mathrm{MAX})^2\right)\left(\sigma_{s_s}^2+\sigma_{s_0}^2+(0.03\times\mathrm{MAX})^2\right)}, \quad (30)$$

where: MAX is the maximum amplitude of ECG signal; $s_{si}$ denotes the i-th element of $\mathbf{s}_s$; $s_{0i}$ denotes the i-th element of $\mathbf{s}_0$; $\mu_{ss}$ denotes the element-wised mean of $\mathbf{s}_s$; $\mu_{s0}$ denotes the element-wised mean of $\mathbf{s}_0$; $\sigma_{ss}$ denotes the element-wised standard deviation of $\mathbf{s}_s$; $\sigma_{s0}$ denotes the element-wised standard deviation of $\mathbf{s}_0$; $\sigma_{sss0}$ denotes the element-wised covariance of $\mathbf{s}_s$ and $\mathbf{s}_0$; dB is the abbreviation of decibel.

### 4.4 Experimental Platforms

The experimental platform contains both a hardware platform and a software platform. The experimental hardware platform comprises an Intel Core CPU and an Nvidia A100 GPU. The experimental software platform includes a Colab client and a server, running on Windows and Linux operating systems respectively.

### 4.5 Experimental Categories

Four experiments are designed to evaluate the performance of the proposed method. The first experiment, focused on performance comparison, assesses the reconstruction quality of the classical and proposed open-loop algorithms (CAESR, DCAESR, and TFSR) and the proposed closed-loop algorithms (CCAESR, CDCAESR, and CTFSR). The second experiment, concentrated on down-sampling methods, compares the reconstruction performance of the interpolation approaches employed in the LR module. The third experiment, centered on hyperparameter values, investigates the impact of the proposed method's hyperparameter on reconstruction performance. The aforementioned three experiments are based on CECGSR architecture I. The fourth experiment, committed to negative feedback, evaluates the reconstruction performance of CECGSR architectures I and II.

### 4.6 Experimental Results

4.6.1 Experimental results of performance comparison

The experimental results of CAESR and CCAESR on the PTB-XL dataset are shown in Table 4. The upper half lists the laboratory results on the original PTB-XL dataset and its five subsets; The lower half lists the laboratory results on the noisy PTB-XL dataset and its five subsets. ΔPSNR is the average PSNR increment of CCAESR compared with CAESR; ΔSSIM is the average SSIM increment of CCAESR compared with CAESR; $\Delta PSNR_m$ is the maximum PSNR increment; $\Delta SSIM_m$ is the maximum SSIM increment. The experimental results indicate that the proposed CCAESR outperforms classical CAESR in signal reconstruction performance metrics, PSNR and SSIM. The experimental results also indicate that the performance metrics on the original PTB-XL dataset are superior to those on the noisy PTB-XL dataset. The experimental results further indicate that the performance metrics on CD subset achieve the maximal PSNR increment and positive SSIM increment.

The laboratory data of DCAESR and CDCAESR on the PTB-XL dataset are displayed in Table 5. The laboratory data illustrate that the proposed CDCAESR surpasses classical DCAESR in signal recovery performance metrics, PSNR and SSIM. The laboratory data also illustrate that the performance metrics on the original PTB-XL dataset outweigh those on the noisy PTB-XL dataset. The laboratory data further illustrate that the performance metrics on MI subset attain the maximal PSNR increment and positive SSIM increment.

The simulation data of TFSR and CTFSR on the PTB-XL dataset are exhibited in Table 6. The simulation data demonstrate that the proposed CTFSR exceeds proposed TFSR in signal restoration performance metrics, PSNR and SSIM. The simulation data also demonstrate that the performance metrics on the original PTB-XL dataset outstrip those on the noisy PTB-XL dataset. The simulation data further demonstrate that the performance metrics on NORM subset gain the maximal PSNR increment and positive SSIM increment.

The performance metrics in Table 5 hold superiority over their counterparts in Table 4. This is due to the fact that DCAE is more suitable than CAE for ECGSR. The performance metrics in Table 6 hold superiority over their counterparts in Table 5. This is owing to the fact that TF is more efficient than DCAE for ECGSR.

**Table 4. Experimental results of CAESR and CCAESR on the PTB-XL dataset.**

| Dataset | Subset | Algorithm | PSNR↑ | ΔPSNR↑ | $\Delta PSNR_m$↑ | SSIM↑ | ΔSSIM↑ | $\Delta SSIM_m$↑ |
|---|---|---|---|---|---|---|---|---|
| Noiseless PTB-XL | CD | CAESR | 23.2278 | **0.2987** | **1.3822** | 0.9890 | **0.0010** | **0.0175** |
| | | CCAESR | 23.5625 | | | 0.9900 | | |
| | HYB | CAESR | 20.4514 | 0.1929 | 0.8972 | 0.9868 | 0.0010 | 0.0045 |
| | | CCAESR | 20.6443 | | | 0.9878 | | |
| | MI | CAESR | 24.7708 | 0.2465 | 1.2383 | 0.9941 | 0.0005 | 0.0050 |
| | | CCAESR | 25.0173 | | | 0.9946 | | |
| | NORM | CAESR | 25.2998 | 0.1995 | 1.1126 | 0.9948 | 0.0004 | 0.0070 |
| | | CCAESR | 25.4993 | | | 0.9952 | | |
| | STTC | CAESR | 24.7417 | 0.1544 | 0.8426 | 0.9937 | 0.0004 | 0.0029 |
| | | CCAESR | 24.9061 | | | 0.9941 | | |
| | Average | CAESR | 23.6983 | 0.2276 | 1.2171 | 0.9917 | 0.0006 | 0.0070 |
| | | CCAESR | 23.9259 | | | 0.9923 | | |
| Noisy PTB-XL | CD | CAESR | 16.3799 | **0.0026** | 0.0374 | 0.9391 | **0.0002** | 0.0004 |
| | | CCAESR | 16.3825 | | | 0.9393 | | |
| | HYB | CAESR | 14.9383 | 0.0007 | **0.0487** | 0.9385 | 0.0001 | 0.0003 |
| | | CCAESR | 14.9390 | | | 0.9386 | | |
| | MI | CAESR | 16.7090 | 0.0006 | 0.0373 | 0.9409 | 0.0001 | 0.0002 |
| | | CCAESR | 16.7096 | | | 0.9410 | | |
| | NORM | CAESR | 17.2476 | 0.0001 | 0.0029 | 0.9403 | 0.0001 | 0.0002 |
| | | CCAESR | 17.2477 | | | 0.9404 | | |
| | STTC | CAESR | 16.7989 | 0.0007 | 0.0341 | 0.9041 | 0.0002 | **0.0009** |
| | | CCAESR | 16.7996 | | | 0.9043 | | |
| | Average | CAESR | 16.4147 | 0.0009 | 0.0221 | 0.9398 | 0.0001 | 0.0003 |
| | | CCAESR | 16.4156 | | | 0.9399 | | |

**Table 5. Laboratory results of DCAESR and CDCAESR on the PTB-XL dataset.**

| Dataset | Subset | Algorithm | PSNR↑ | ΔPSNR↑ | $\Delta PSNR_m$↑ | SSIM↑ | ΔSSIM↑ | $\Delta SSIM_m$↑ |
|---|---|---|---|---|---|---|---|---|
| Noiseless PTB-XL | CD | DCAESR | 23.1463 | **0.3817** | 1.6294 | 0.9891 | **0.0008** | 0.0052 |
| | | CDCAESR | 23.5280 | | | 0.9899 | | |
| | HYB | DCAESR | 20.9985 | 0.1678 | 0.9926 | 0.9881 | 0.0006 | 0.0034 |
| | | CDCAESR | 21.1663 | | | 0.9887 | | |
| | MI | DCAESR | 24.9347 | 0.3443 | **1.6864** | 0.9946 | 0.0004 | 0.0045 |
| | | CDCAESR | 25.2817 | | | 0.9950 | | |
| | NORM | DCAESR | 26.5967 | 0.2474 | 0.9363 | 0.9966 | 0.0002 | **0.0057** |
| | | CDCAESR | 26.8441 | | | 0.9968 | | |
| | STTC | DCAESR | 25.7333 | 0.2279 | 0.9124 | 0.9953 | 0.0002 | 0.0028 |
| | | CDCAESR | 25.9612 | | | 0.9955 | | |
| | Average | DCAESR | 24.2824 | 0.2738 | 1.2041 | 0.9927 | 0.0004 | 0.0048 |
| | | CDCAESR | 24.5562 | | | 0.9931 | | |
| Noisy | CD | DCAESR | 22.9972 | **0.3440** | 1.3725 | 0.9889 | **0.0008** | **0.0058** |
| PTB-XL | | CDCAESR | 23.3412 | | | 0.9897 | | |
| | HYB | DCAESR | 20.9030 | 0.1519 | 0.9474 | 0.9880 | 0.0005 | 0.0038 |
| | | CDCAESR | 21.0549 | | | 0.9885 | | |
| | MI | DCAESR | 24.7584 | 0.2987 | **1.5185** | 0.9944 | 0.0004 | 0.0044 |
| | | CDCAESR | 25.0571 | | | 0.9948 | | |
| | NORM | DCAESR | 26.3904 | 0.2124 | 0.9205 | 0.9964 | 0.0002 | 0.0055 |
| | | CDCAESR | 26.6028 | | | 0.9966 | | |
| | STTC | DCAESR | 25.5718 | 0.1973 | 0.9110 | 0.9952 | 0.0002 | 0.0029 |
| | | CDCAESR | 25.7691 | | | 0.9954 | | |
| | Average | DCAESR | 24.1242 | 0.2408 | 1.1228 | 0.9926 | 0.0004 | 0.0048 |
| | | CDCAESR | 24.3650 | | | 0.9930 | | |

**Table 6. Simulation results of TFSR and CTFSR on the PTB-XL dataset.**

| Dataset | Subset | Algorithm | PSNR↑ | ΔPSNR↑ | $\Delta PSNR_m$↑ | SSIM↑ | ΔSSIM↑ | $\Delta SSIM_m$↑ |
|---|---|---|---|---|---|---|---|---|
| Noiseless PTB-XL | CD | TFSR | 37.0290 | 0.0234 | 0.3160 | 0.9472 | 0.0000 | **0.0015** |
| | | CTFSR | 37.0524 | | | 0.9472 | | |
| | HYB | TFSR | 37.8945 | **0.0332** | 0.2373 | 0.9608 | **0.0001** | 0.0002 |
| | | CTFSR | 37.9277 | | | 0.9609 | | |
| | MI | TFSR | 36.9492 | 0.0265 | **0.8587** | 0.9490 | 0.0000 | 0.0014 |
| | | CTFSR | 36.9757 | | | 0.9490 | | |
| | NORM | TFSR | 38.2528 | 0.0299 | 0.4230 | 0.9582 | 0.0000 | 0.0003 |
| | | CTFSR | 38.2827 | | | 0.9582 | | |
| | STTC | TFSR | 37.5589 | 0.0317 | 0.3818 | 0.9493 | 0.0001 | 0.0002 |
| | | CTFSR | 37.5906 | | | 0.9494 | | |
| | Average | TFSR | 37.6737 | 0.0287 | 0.5047 | 0.9536 | 0.0000 | 0.0006 |
| | | CTFSR | 37.7024 | | | 0.9536 | | |
| Noisy PTB-XL | CD | TFSR | 34.2103 | **0.0408** | 1.3620 | 0.9121 | **0.0005** | 0.0010 |
| | | CTFSR | 34.2510 | | | 0.9126 | | |
| | HYB | TFSR | 35.5270 | 0.0342 | 0.7385 | 0.9331 | 0.0003 | 0.0027 |
| | | CTFSR | 35.5611 | | | 0.9334 | | |
| | MI | TFSR | 34.2950 | 0.0394 | 2.0538 | 0.9134 | 0.0004 | 0.0027 |
| | | CTFSR | 34.3343 | | | 0.9138 | | |
| | NORM | TFSR | 36.0563 | 0.0334 | **5.9317** | 0.9344 | 0.0002 | **0.0046** |
| | | CTFSR | 36.0897 | | | 0.9346 | | |
| | STTC | TFSR | 35.2835 | 0.0324 | 0.6640 | 0.9201 | 0.0002 | 0.0014 |
| | | CTFSR | 35.3159 | | | 0.9204 | | |
| | Average | TFSR | 35.2711 | 0.0357 | 3.4484 | 0.9247 | 0.0003 | 0.0031 |
| | | CTFSR | 35.3067 | | | 0.9249 | | |

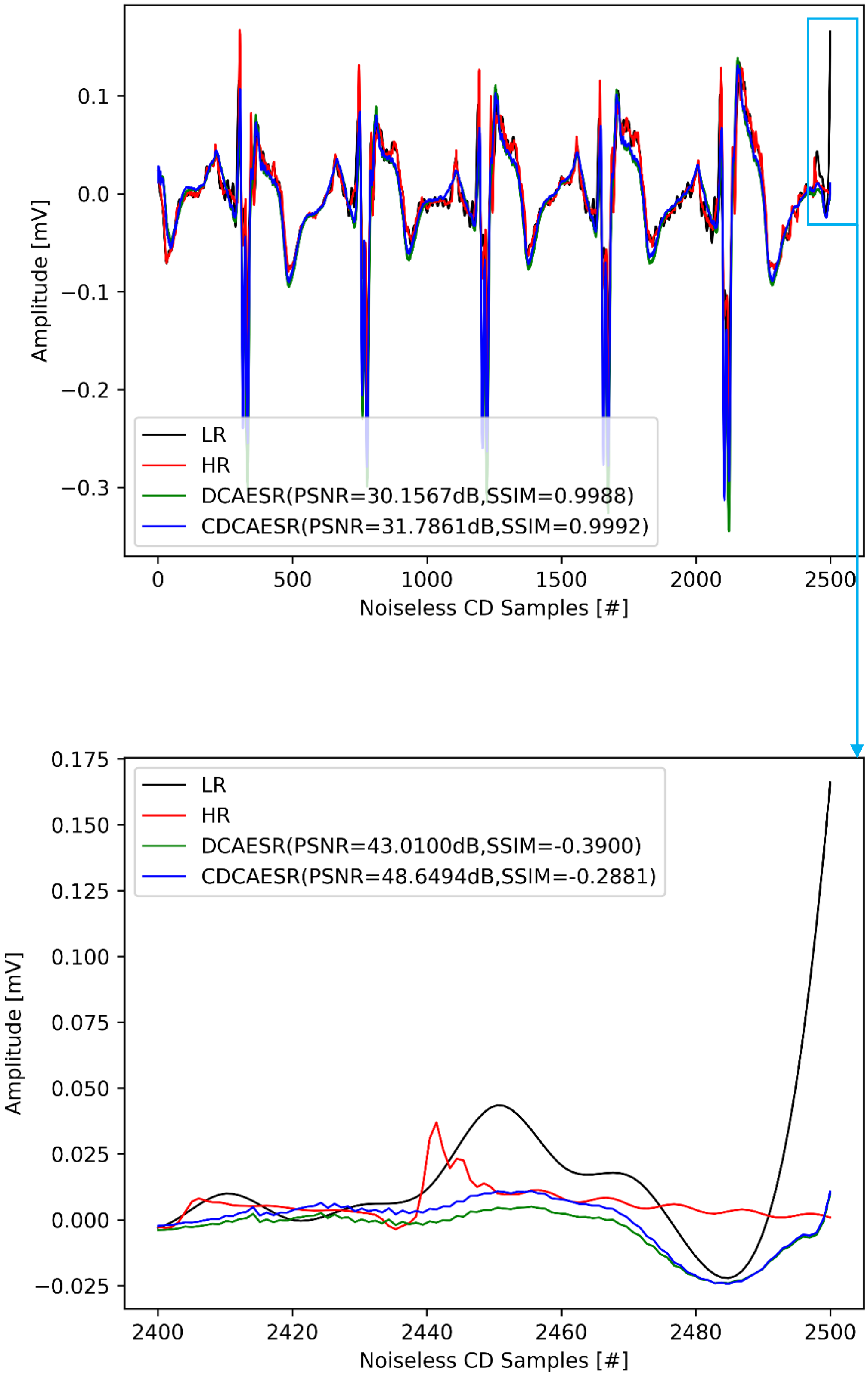

**Figure 7. Experimental results of the original CD subset. The bottom-half is the local enlargement of the related top-half.**

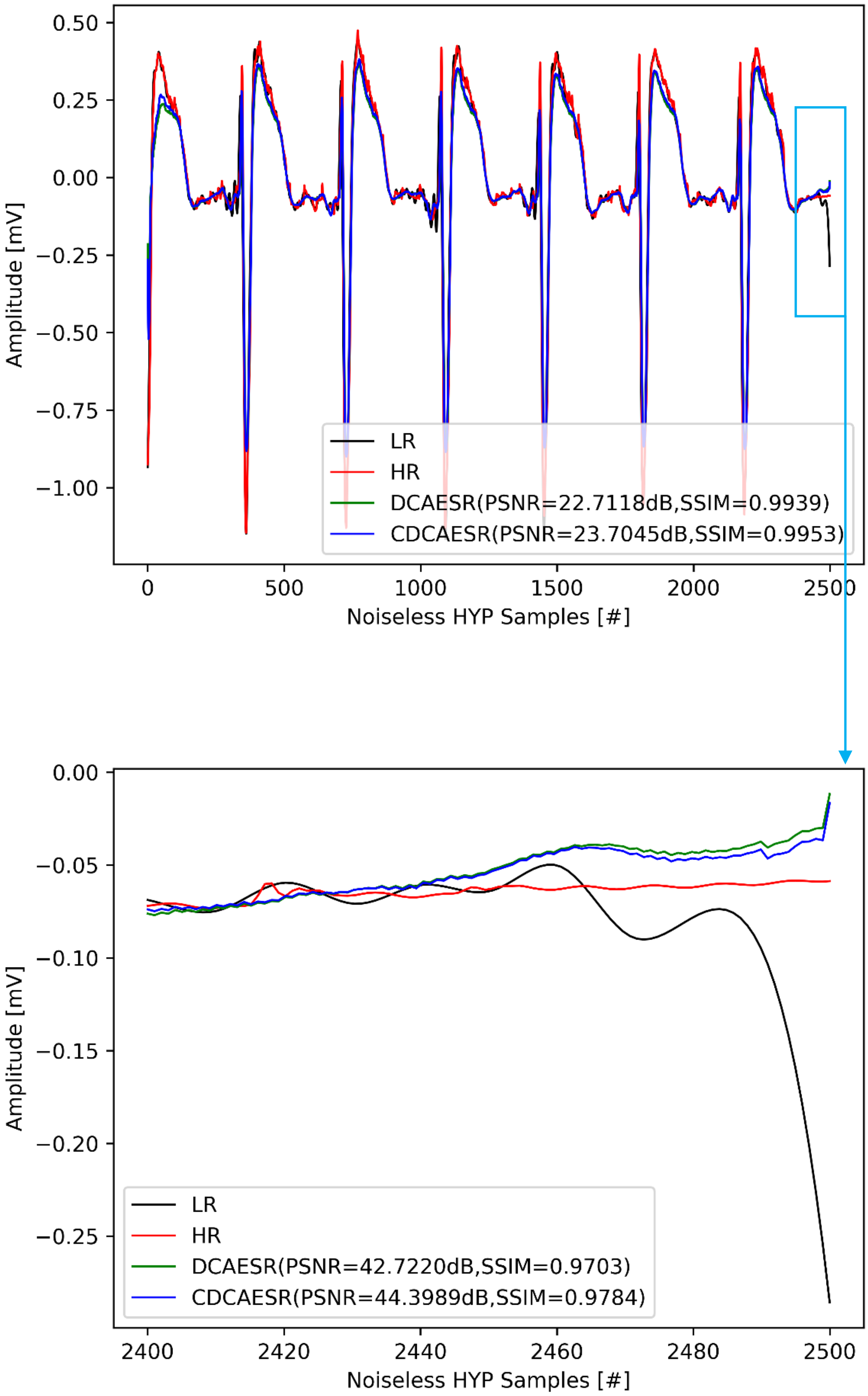

**Figure 8. Experimental results of the original HYP subset. The bottom-half is the local enlargement of the related top-half.**

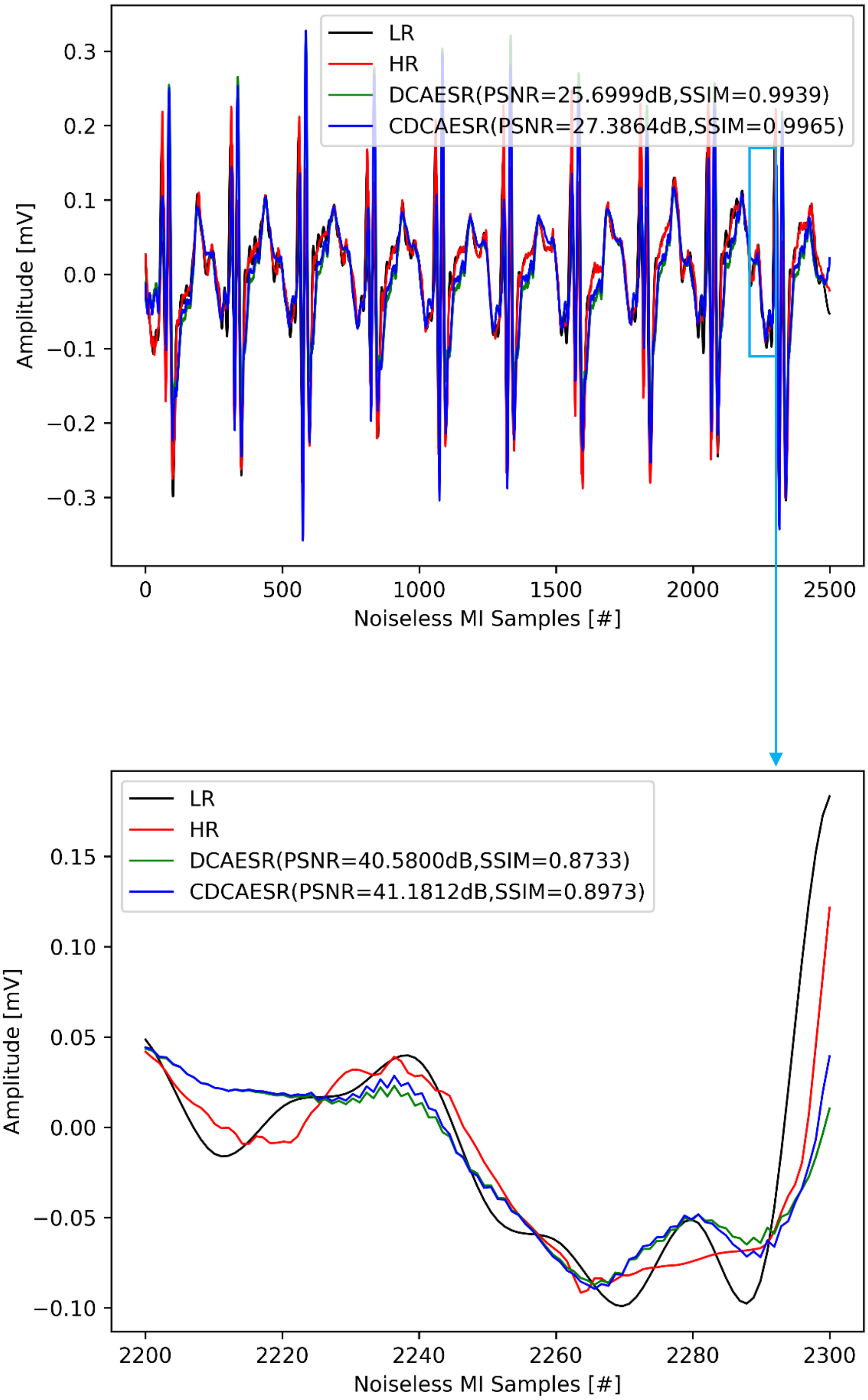

**Figure 9. Experimental results of the original MI subset. The bottom-half is the local enlargement of the related top-half.**

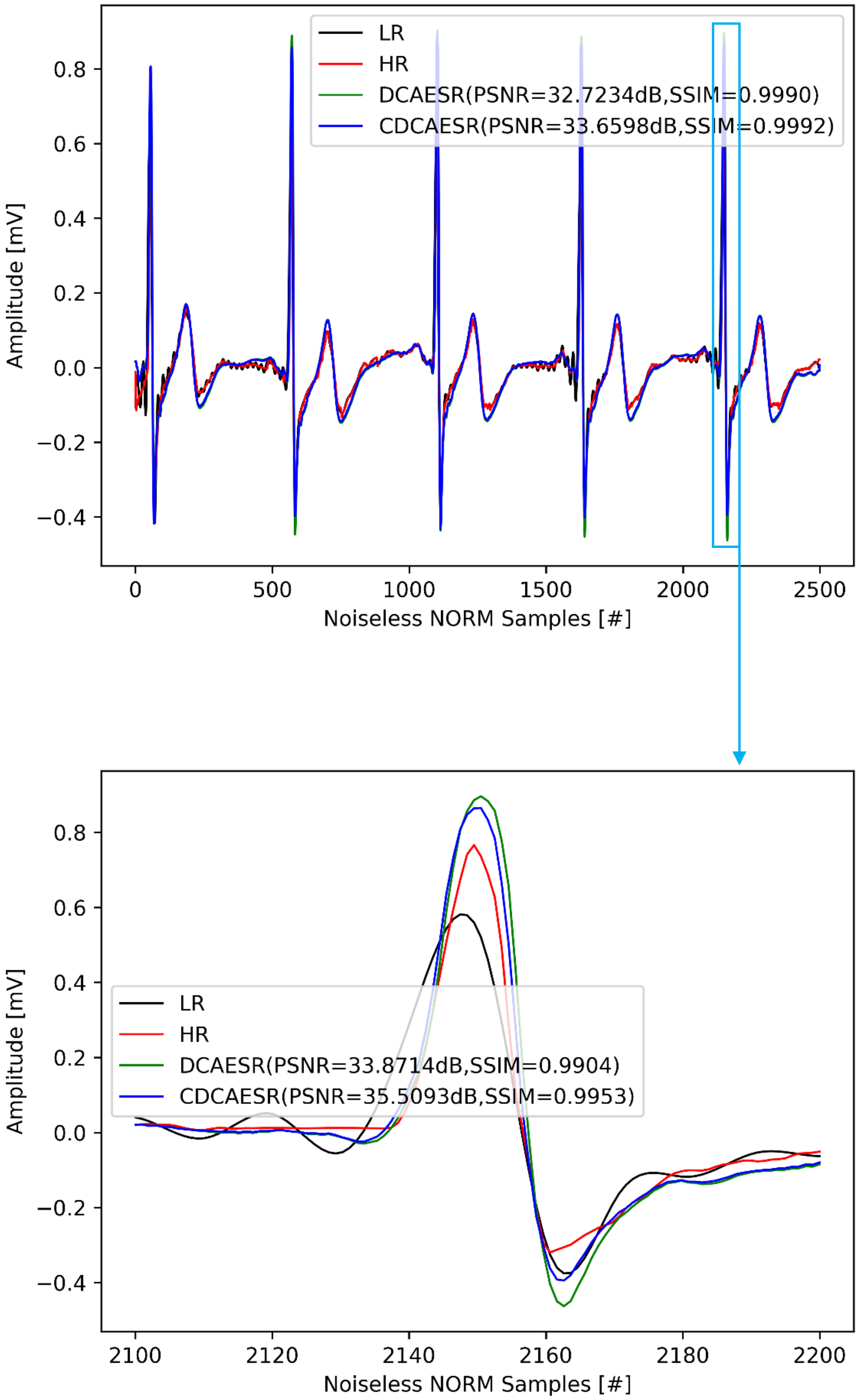

**Figure 10. Experimental results of the original NORM subset. The bottom-half is the local enlargement of the related top-half.**

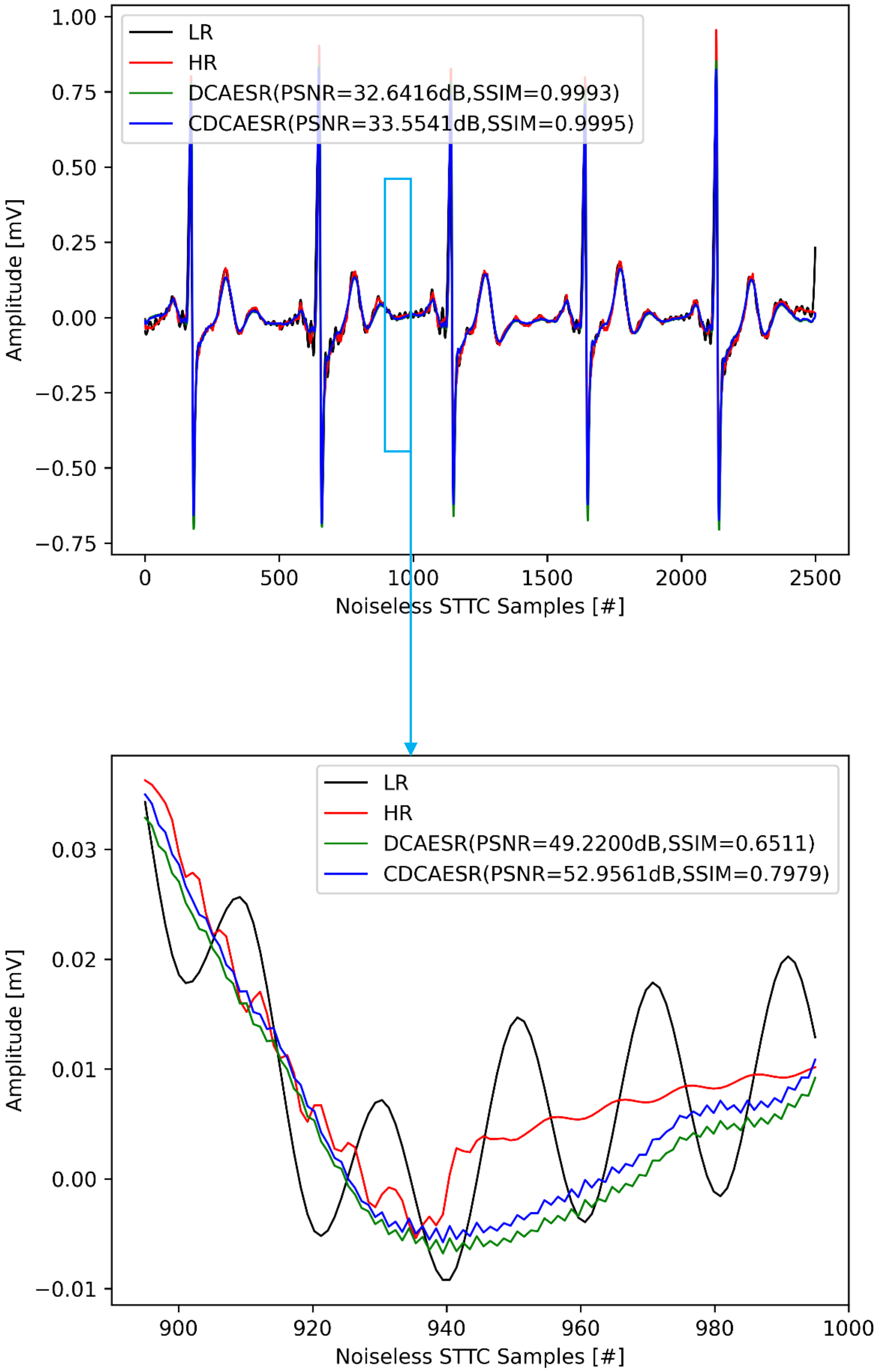

**Figure 11. Experimental results of the original STTC subset. The bottom-half is the local enlargement of the related top-half.**

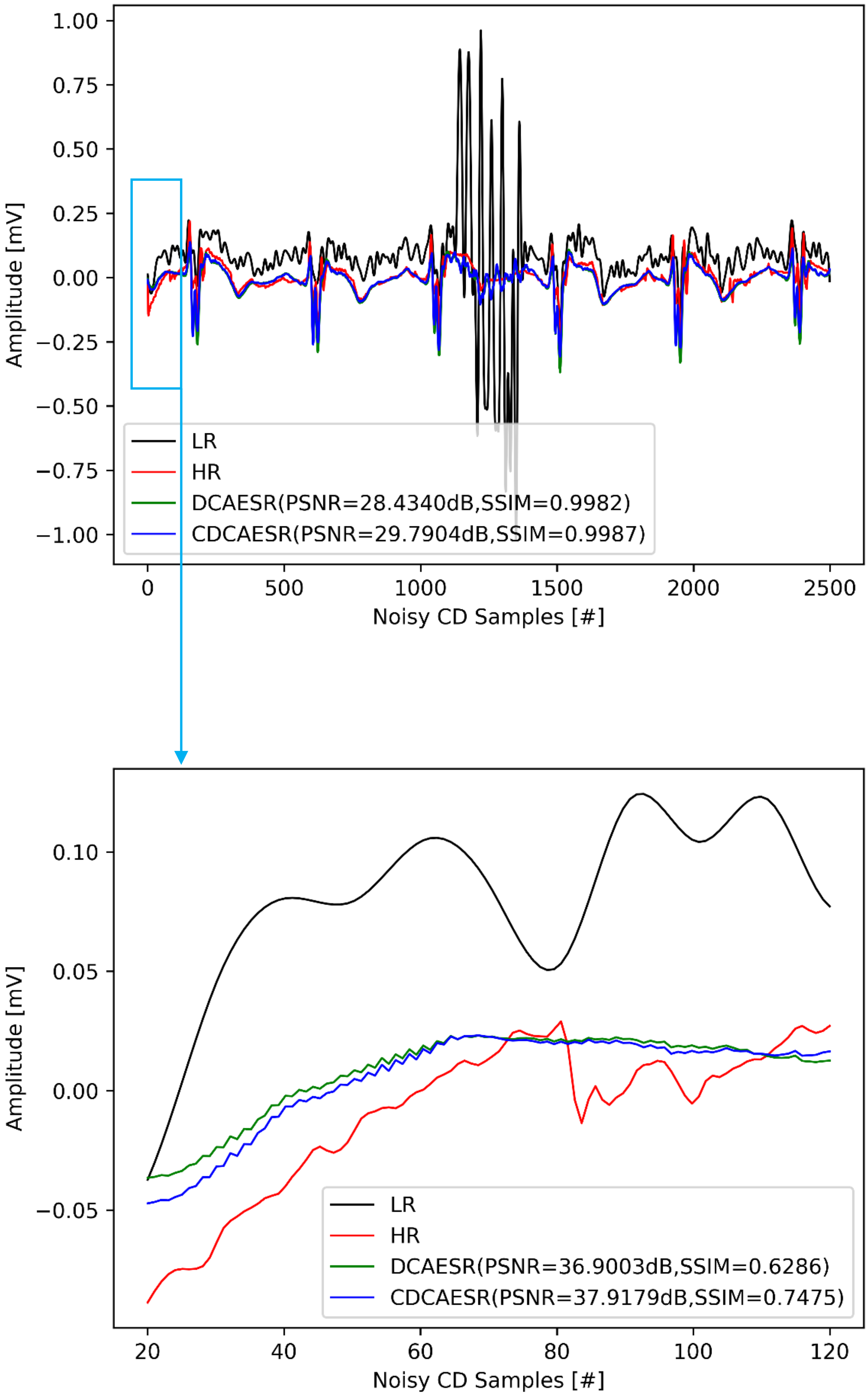

**Figure 12. Experimental results of the noisy CD subset. The bottom-half is the local enlargement of the related top-half.**

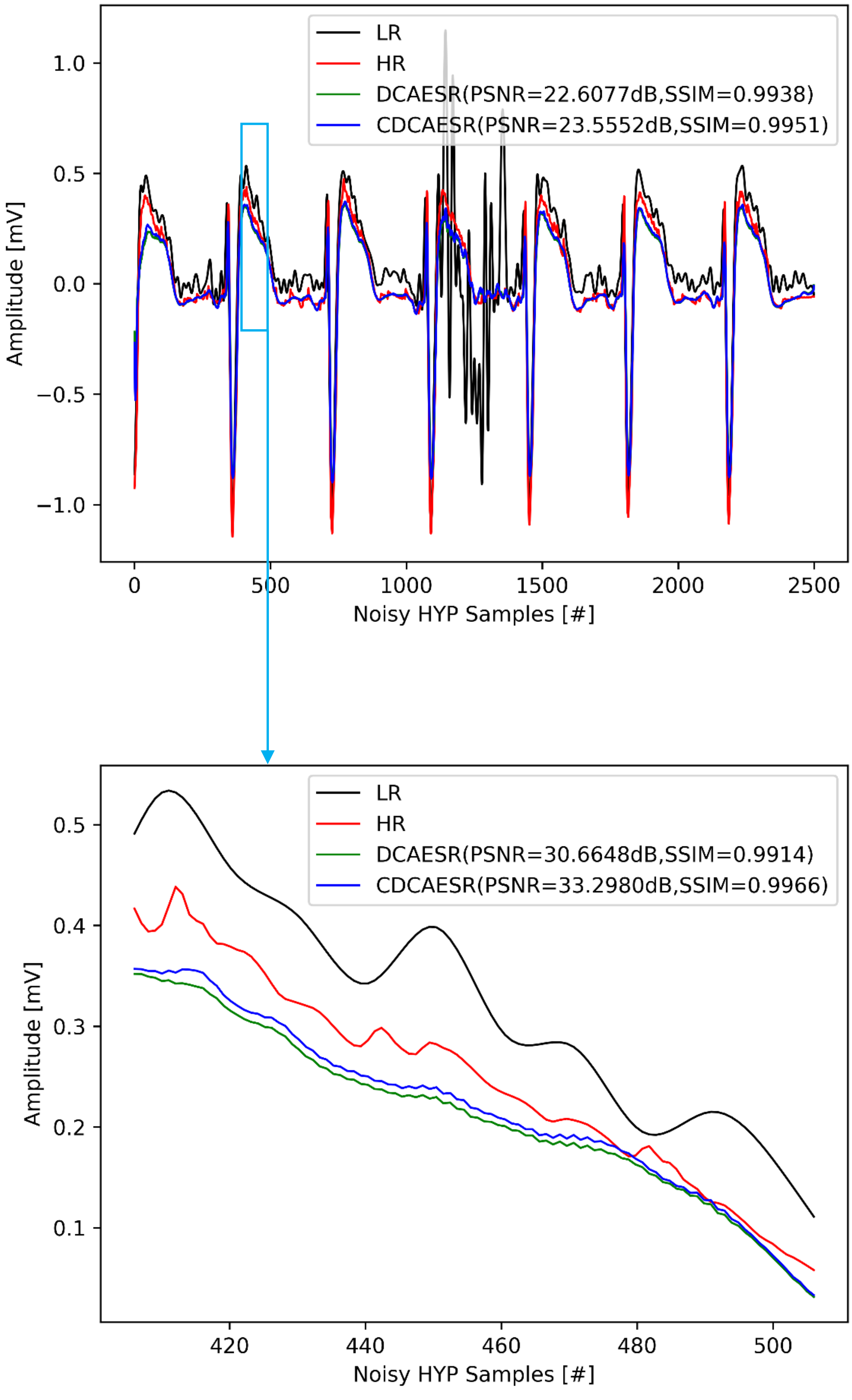

**Figure 13. Experimental results of the noisy HYP subset. The bottom-half is the local enlargement of the related top-half.**

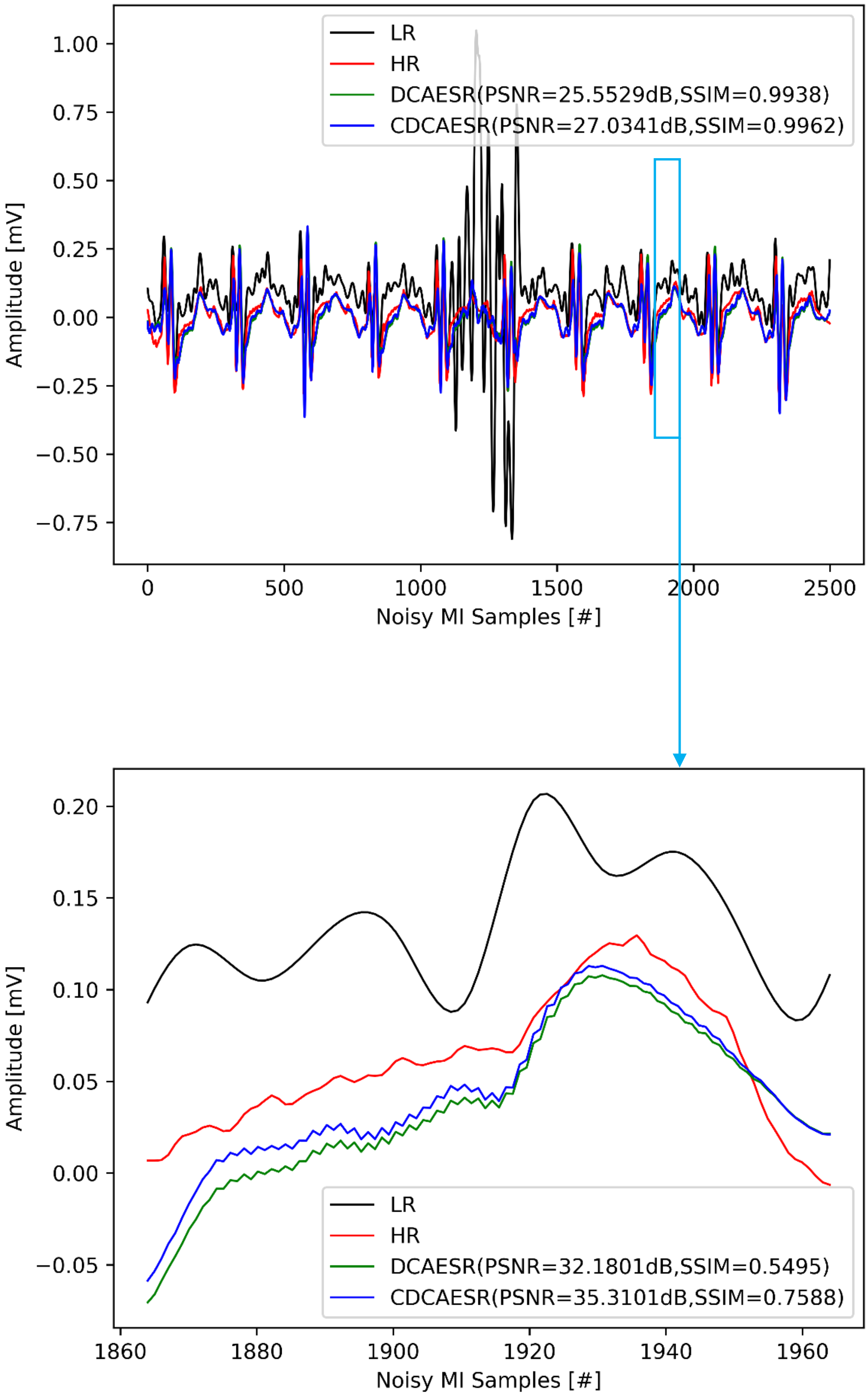

**Figure 14. Experimental results of the noisy MI subset. The bottom-half is the local enlargement of the related top-half.**

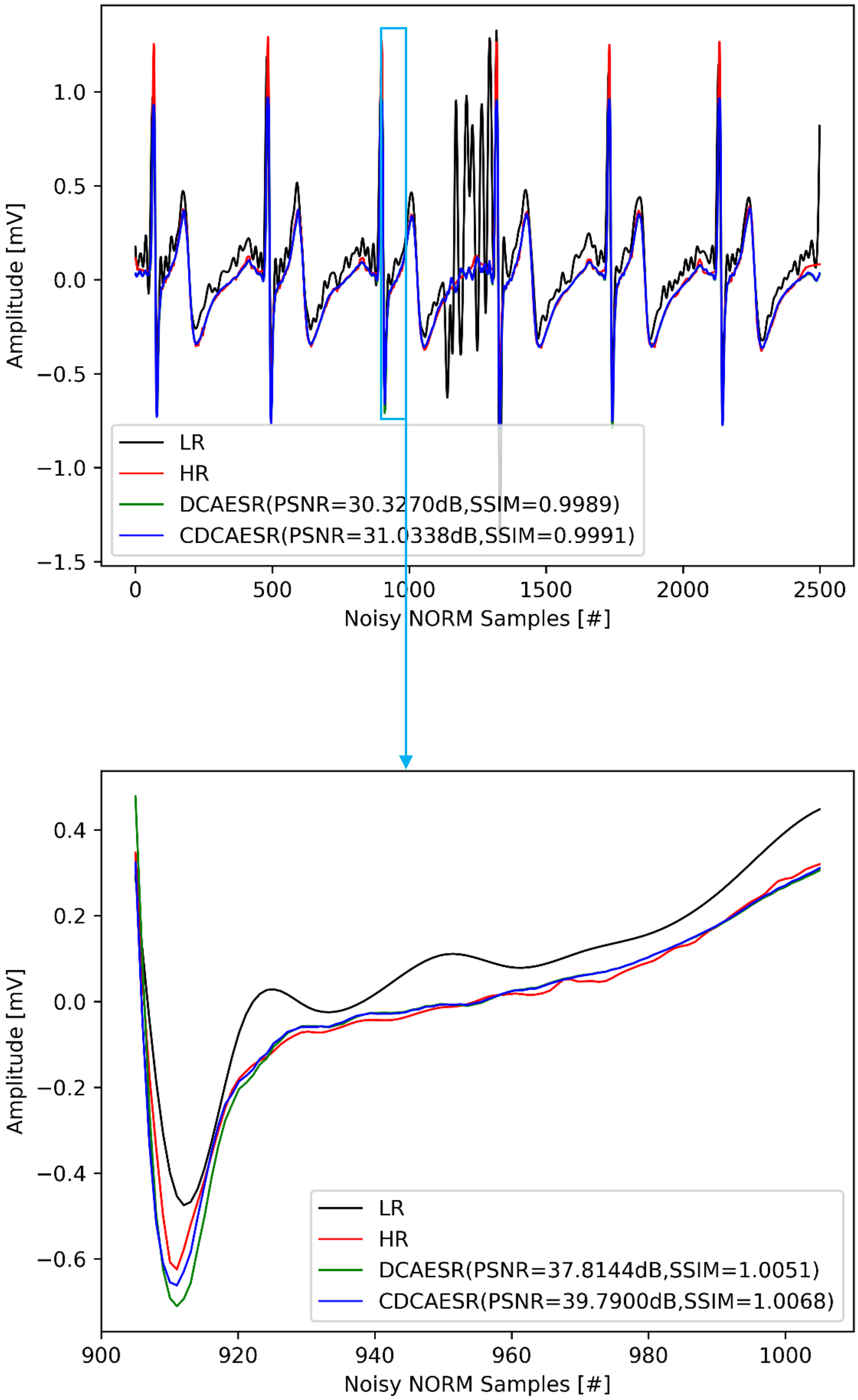

**Figure 15. Experimental results of the noisy NORM subset. The bottom-half is the local enlargement of the related top-half.**

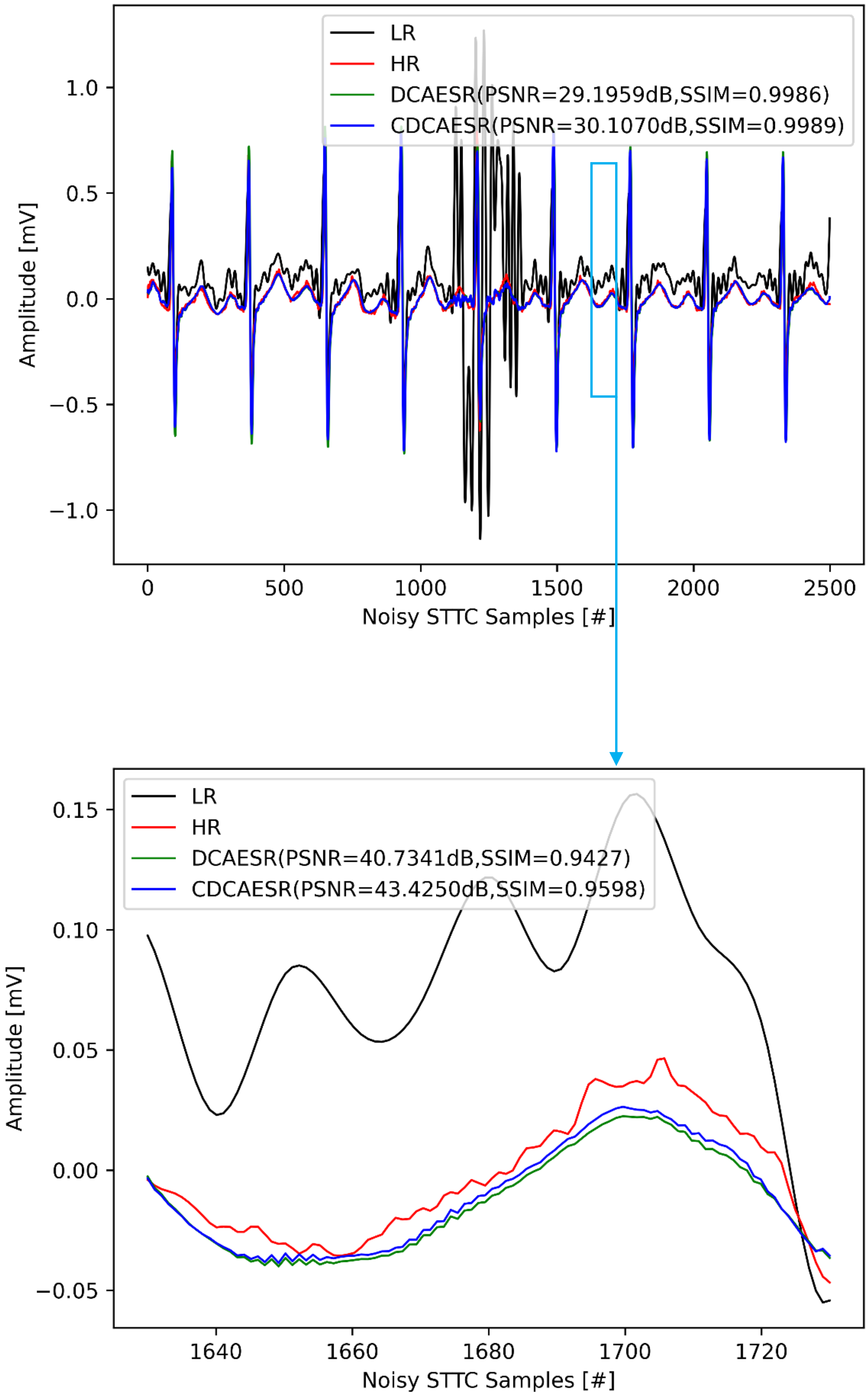

**Figure 16. Experimental results of the noisy STTC subset. The bottom-half is the local enlargement of the related top-half.**

To vividly compare the reconstruction performance of DCAESR and CDCAESR on the original PTB-XL dataset, Figures 8 to 12 respectively exhibit the ECG signals of the five subsets. The bottom-half of Figure 8 is the local enlargement of the related top-half; LR marked with black line denotes the LR signal and is up-sampled to the same size as the HR signal by nearest-neighbor interpolation; HR marked with red line denotes the true SR signal; DCAESR with green line denotes the SR signal of DCAESR; CDCAESR marked with blue line denotes the SR signal of CDCAESR.

Figures 9 to 12 are similar to Figure 8. The experimental results demonstrate that the CDCAESR signal is closer to the HR signal than the DCAESR signal and the LR signal. The experimental results also demonstrate that the reconstruction performance metrics, PSNR and SSIM, of the CDCAESR is higher than those of DCAESR. It should be mentioned that DCAESR and CDCAESR in the bottom half of Figure 8 have minus SSIM due to the huge structural difference between the reconstruction SR signal of DCAESR or CDCAESR and the true SR signal.

To graphically compare the reconstruction performance of DCAESR and CDCAESR on the noisy PTB-XL dataset, Figures 13 to 17 respectively expose the ECG signals of the five noisy subsets. Figures 13 to 17 are analogous to Figures 8 to 12. The laboratory data reveal that the gap between the CDCAESR signal and the true SR signal is smaller than that between the DCAESR signal and the true SR signal and that between the LR signal and the true SR signal. The laboratory data also reveal that the reconstruction performance metrics, PSNR and SSIM, of CDCAESR is higher than those of DCAESR.

In short, the proposed closed-loop CECGSR outperforms the classical open-loop ECGSR in ECG signal reconstruction performance.

4.6.2 Experimental results of down-sampling approaches

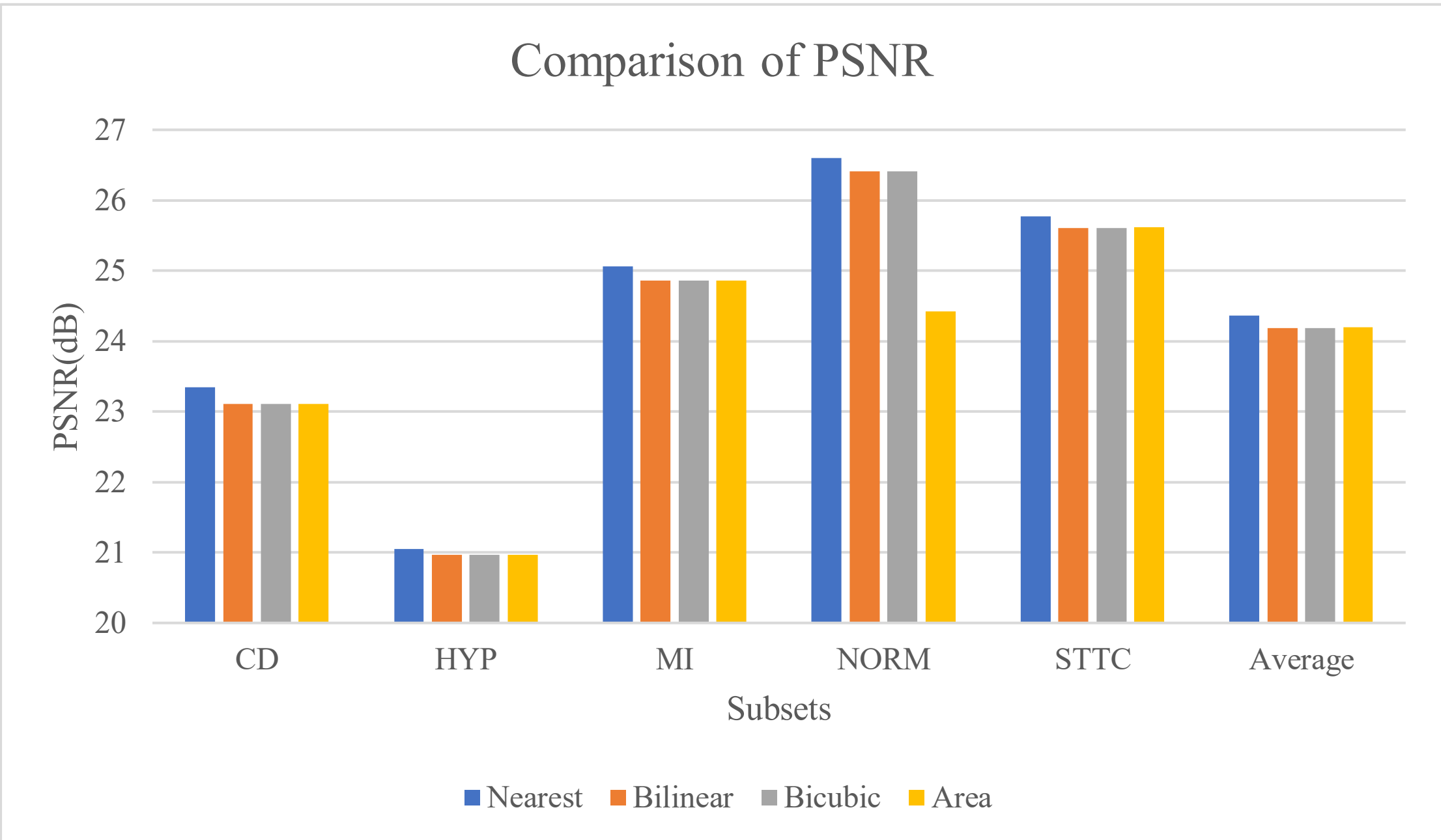

**Figure 17. Comparison of PSNR for different down-sampling interpolation approaches.**

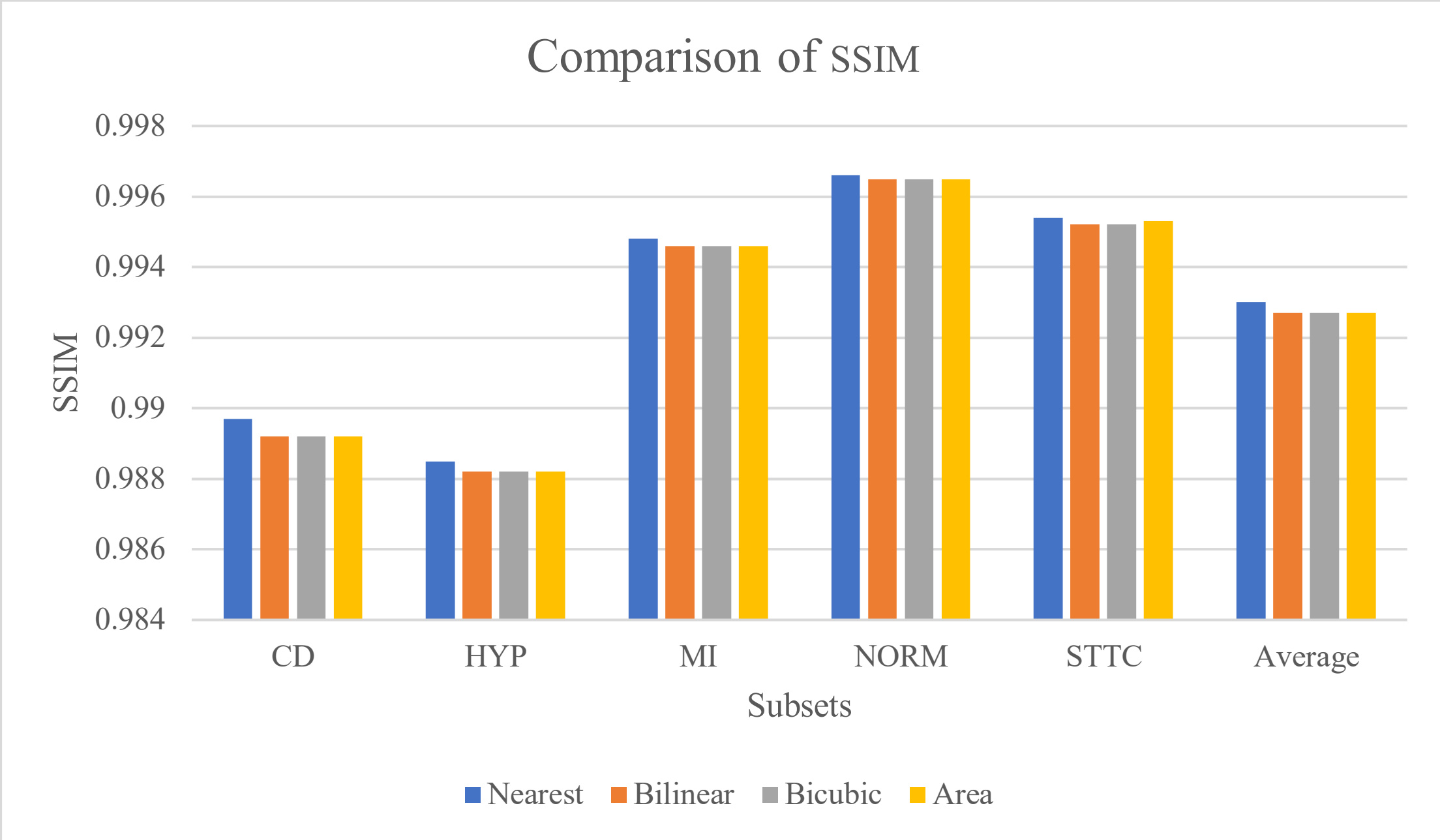

**Figure 18. Comparison of SSIM for different down-sampling interpolation approaches.**

The comparisons of performance metrics, PSNR and SSIM, for the four down-sampling approaches, nearest-neighbor interpolation, bilinear interpolation, bicubic interpolation, and area interpolation, utilized in LR module are shown in Figures 17 and 18. The experimental results come from CDCAESR on the noisy PTB-XL dataset. The experimental results disclose that the nearest-neighbor interpolation holds the best performance.

4.6.3 Experimental results of hyperparameter values

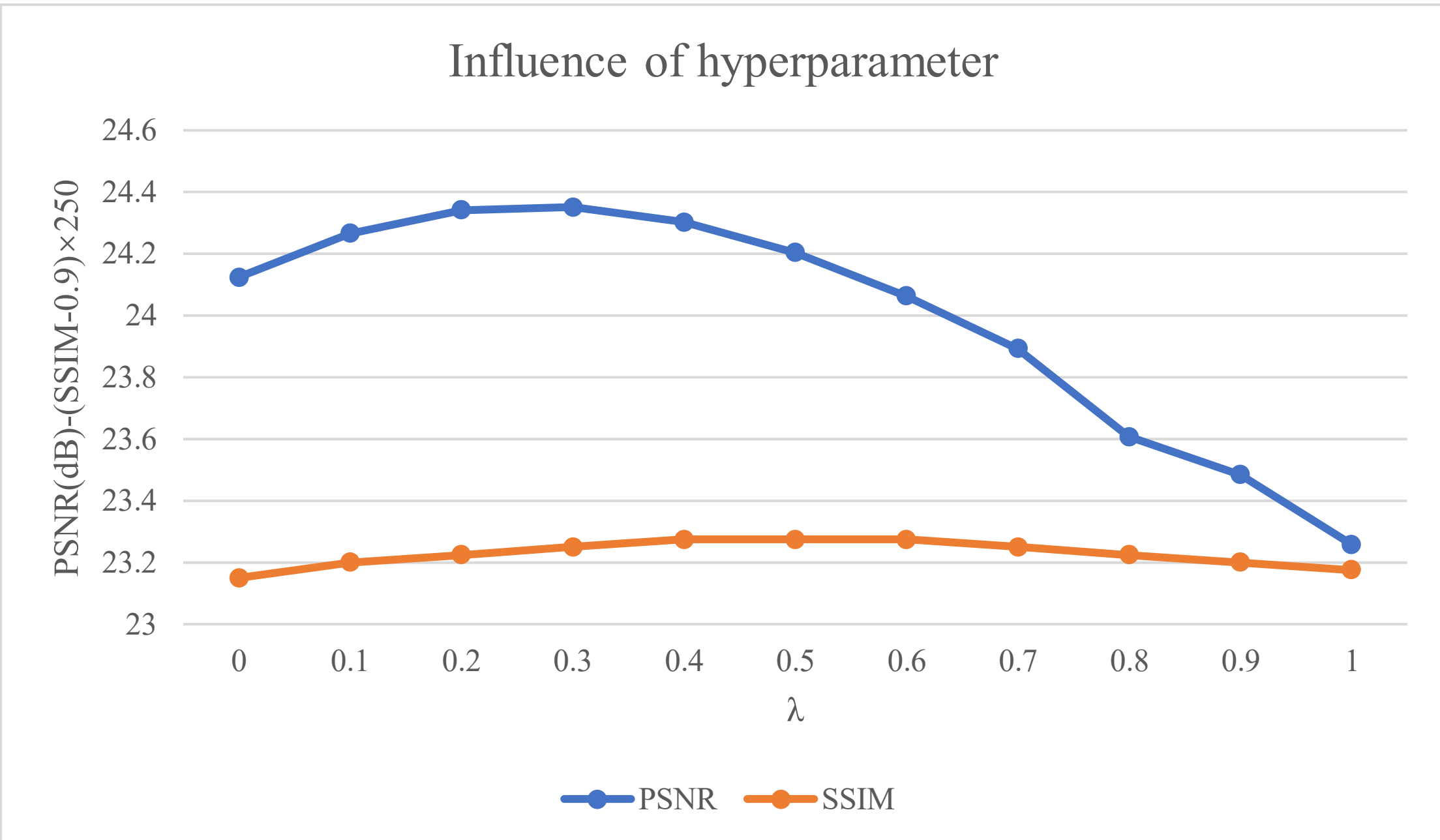

**Figure 19. Influence of hyperparameter.**

The influence of hyperparameter λ in Equation (22) on performance metrics, PSNR and SSIM, is displayed in Figure 19. The laboratory data stem from CDAESR on the noisy PTB-XL dataset. The laboratory data manifest that PSNR reaches the maximum at λ = 0.3; SSIM reaches the maximum at λ = 0.5; they cannot simultaneously reach the maximum at the same λ. The proposed method chooses the optimal λ with the maximal PSNR and a positive SSIM increment.

4.6.4 Experimental results of negative feedback

The experimental results of negative feedback applied to the CDCAESR algorithm on both noiseless and noisy PTB-XL datasets are enumerated in Table 6. The performance metric PSNR of CECGSR architecture I outperforms that of CECGSR architecture II with comparable performance metric SSIM. In other words, for the negative feedback mechanism, the differences between LR samples are greater than those between SR samples in PSNR with comparable SSIM. This is owning to the fact that the initial SR reconstruction $\mathbf{s}_{s0}$ in CECGSR architecture II is inaccurate. It should be noted that the performance metrics in Table 6 represent the average of all testing samples, whereas those in Tables 4 and 5 are the average of five testing subsets.

**Table 6. Experimental results of negative feedback.**

| Algorithm | Dataset | Architecture | PSNR(dB)↑ | SSIM↑ |
|---|---|---|---|---|
| CDCAESR | Noiseless PTB-XL | CECGSR I | **25.6365** | 0.9949 |
| | | CECGSR II | 25.6132 | **0.9956** |
| | Noisy PTB-XL | CECGSR I | **25.4194** | 0.9947 |
| | | CECGSR II | 25.4157 | **0.9955** |

## 5 Conclusions

This paper proposes a closed-loop method, CECGSR, based on the theory of automatic control. The architecture of CECGSR consists of four elements: summator, PP element, SR element, and LR element. The summator introduces negative feedback based on the differences between the LR ECG signals; the PP element utilizes the traditional PI control; the SR element takes advantages any existing or proposed pretrained open-loop ECGSR models based on the PnP strategy; LR element wields nearest-neighbor interpolation. The closed-loop framework is portrayed by the mathematical nonlinear loop equation. That the steady-state error of the proposed infrastructure is nearly close to zero is proven by local approximation of Taylor series expansion. This paper also presents TF model based ECGSR and CECGSR approaches. The laboratory results on the noiseless and noisy PTB-XL datasets show that the proposed method is superior to the state-of-the-art competing algorithms of open-loop ECGSR in reconstruction performance. TF dominates DCAE and CAE in recovery capability. The laboratory results also show that the nearest-neighbor interpolation adopted in LR element possesses the best performance. The laboratory results further show the proposed method is sensitive to the hyperparameter. The proposed method selects an optimal hyperparameter with the maximum PSNR and a positive SSIM increment. The laboratory results finally show that the CECGSR architecture I based on the differences between the LR samples outperforms the CECGSR architecture II in performance metric PSNR with comparable SSIM. The proposed method will significantly increase ECG signal details and decrease ECG signal artifacts in clinical applications.

In our future work, we will investigate more ECG datasets and pretrained open-loop ECGSR

algorithms, explore more sophisticated LR units, and further probe advanced PP modules based on modern control methods such as fuzzy and adaptive control.

# Statements and Declarations

### 1 Funding

The authors declare that no funds, grants, or other support were received during the preparation of this manuscript.

**2 Competing Interests**
The authors have no relevant financial or non-financial interests to disclose.

**3 Author Contributions**
Conceptualization, Honggui LI and Maria TROCAN; Methodology, Honggui LI and Dimitri GALAYKO; Writing, Honggui LI and Zhengyang ZHANG; Experiment, Honggui LI, Dingtai LI, Sinan CHEN, Nahid MD LOKMAN HOSSAIN, Hantao LU, Ruobing WANG, Xinfeng XU, Yinlu QIN, and Yuting FENG; Supervision, Amara AMARA and Mohamad SAWAN. All authors have read and agreed to the published version of the manuscript.

**4 Data Availability**
The datasets analyzed during the current study are available in the internet repositories, https://github.com/UgoLomoio/DCAE-SR and https://physionet.org/content/ptb-xl/1.0.3/.

**5 Ethics Approval**
This research doesn’t involve human or animal subjects, hence no ethical approval is required.

**6 Consent to Participate**
This research doesn’t involve human subjects, hence no consent to participation is required.

**7 Consent to Publish**
This research doesn’t involve human subjects, hence no consent to publication is required.

## Acknowledgment

The authors would very much like to thank all the authors of the competing algorithms for selflessly releasing their source codes of ECGSR on the Microsoft GitHub website. The open-source codes allow us to easily implement the proposed algorithm depending on the competing algorithms. The authors also would very much like to express deep thanks to Google COLAB for its free GPU computing service.

## Biographies

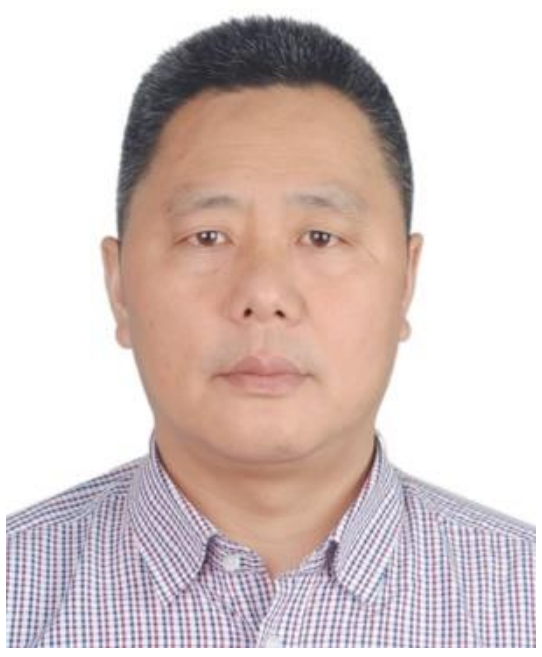

**Honggui LI** received a B.S. degree in electronic science and technology from Yangzhou University and received a Ph.D. degree in mechatronic engineering from Nanjing University of Science and Technology. He is a senior member of the Chinese Institute of Electronics. He is a visiting scholar and a post-doctoral fellow at Institut Supérieur d'Électronique de Paris for one year. He is an associate professor of electronic science and technology and a postgraduate supervisor of electronic science and technology at Yangzhou University. He is a reviewer for some international journals and conferences. He is the author of over 30 refereed journal and conference articles. His current research interests include image processing, machine learning, and integrated circuits engineering.

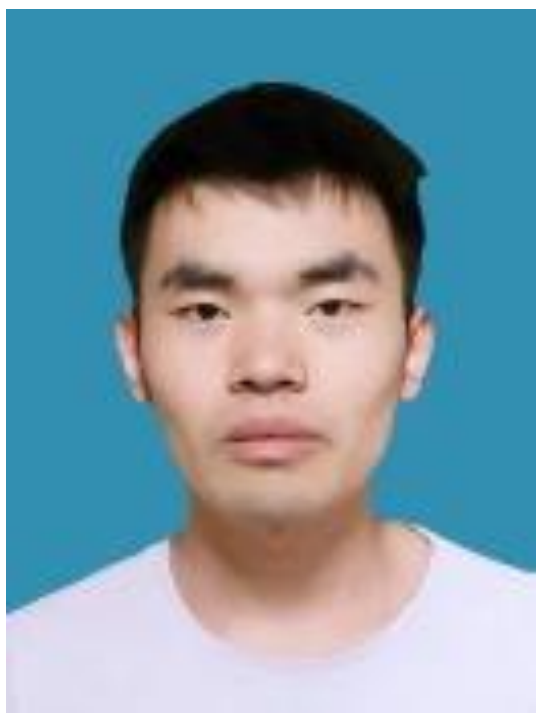

**Zhengyang ZHANG** received a B.E. degree in electronics information engineering from Yangzhou University in China. He is now studying for a master's degree in communication engineering at Yangzhou University in China. His research interests include image super-resolution, machine learning, and communication engineering.

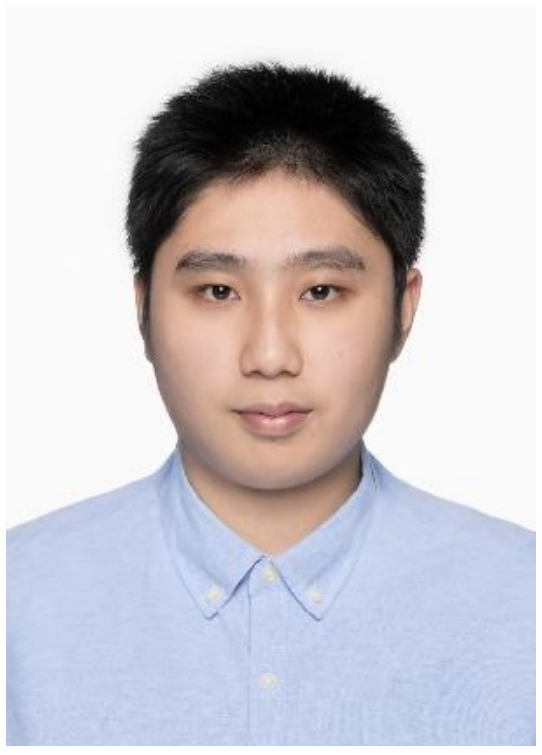

**Dingtai LI** received a B.M. degree in rehabilitation of Chinese medicine from Nanjing University of Chinese Medicine in China. He is now studying for a master's degree in acupuncture and massage at Shanghai University of Traditional Chinese Medicine in China. His research interests include Qigong, acupuncture, massage, image processing, and machine learning.

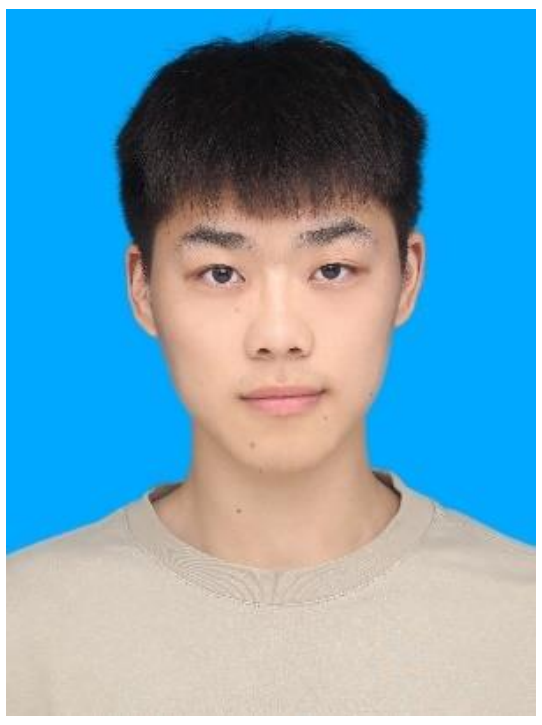

**Sinan CHEN** received a B.E. degree in electronic information engineering from Yangzhou University in China. He is now studying for a master's degree in integrated circuits engineering at Yangzhou University in China. His research interests include image compressive sensing, machine learning, and integrated circuits engineering.

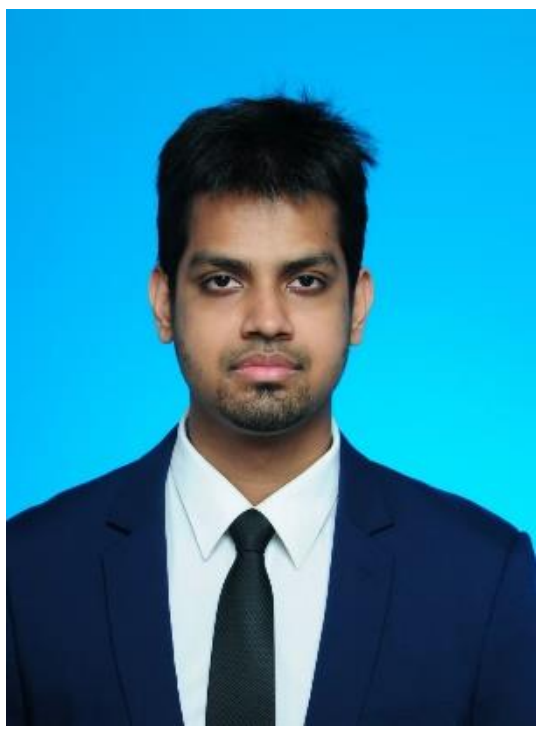

**Nahid MD LOKMAN HOSSAIN** received a B.S. degree in computer science and technology from Chongqing University of Posts and Telecommunications in China. He is now studying for a master's degree in software engineering at Yangzhou University in China. His research interests include image super-resolution, machine learning, innovation software, cloud computing, and big data.

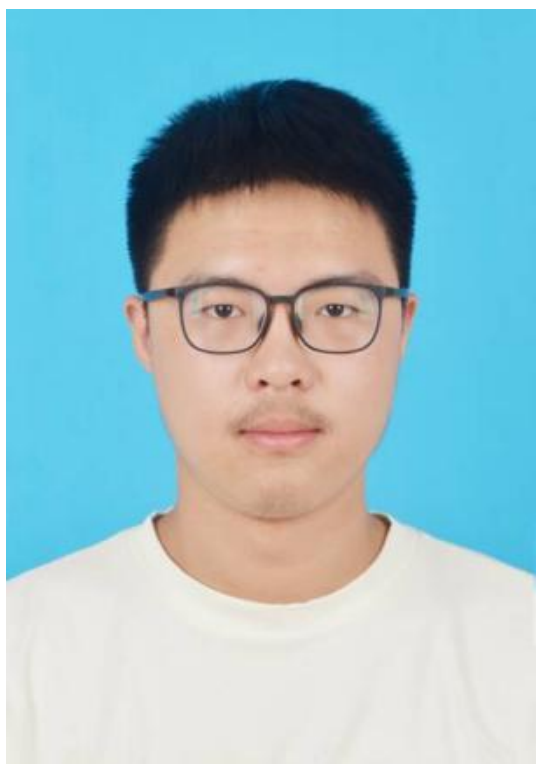

**Hantao LU** is now studying for a B.E. degree in electronic information engineering at Yangzhou University in China. His research interests include image processing and machine learning.

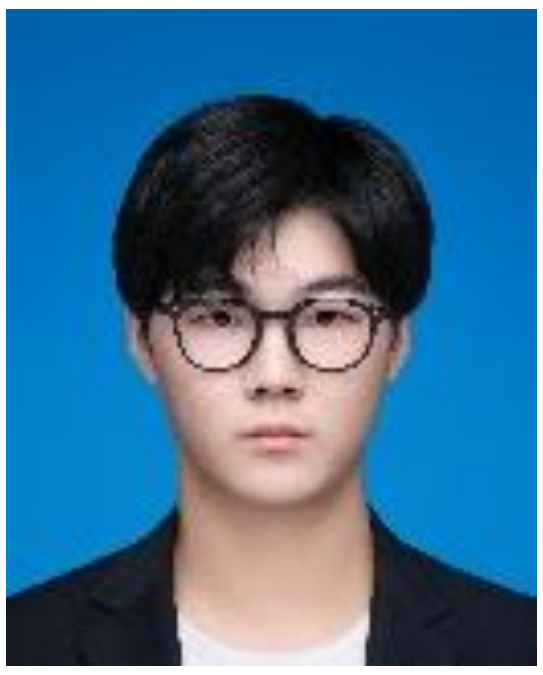

**Ruobing WANG** is now studying for a B.E. degree in electronic information engineering at Yangzhou University in China. His research interests include image processing and machine learning.

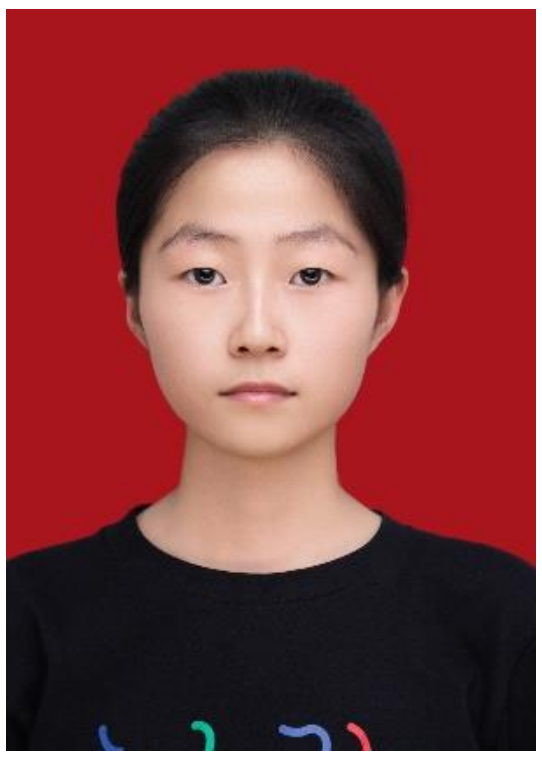

**Xinfeng XU** is now studying for a B.E. degree in electronic information engineering at Yangzhou University in China. Her research interests include image processing and machine learning.

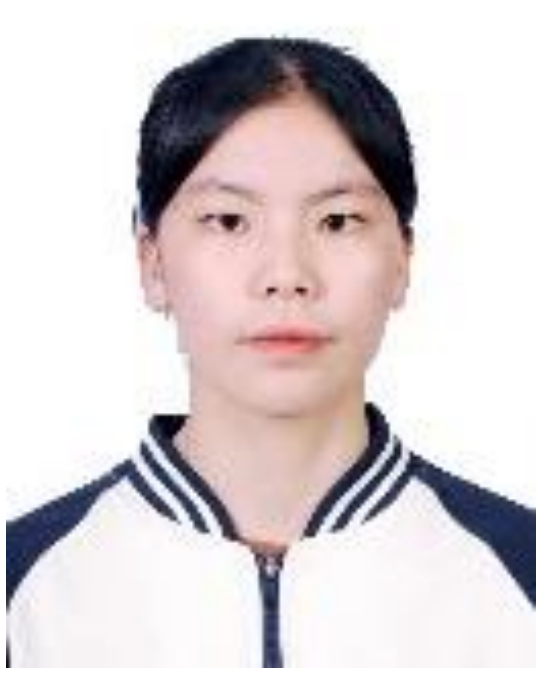

**Yinlu QIN** is now studying for a B.E. degree in electronic information engineering at Yangzhou University in China. Her research interests include image processing and machine learning.

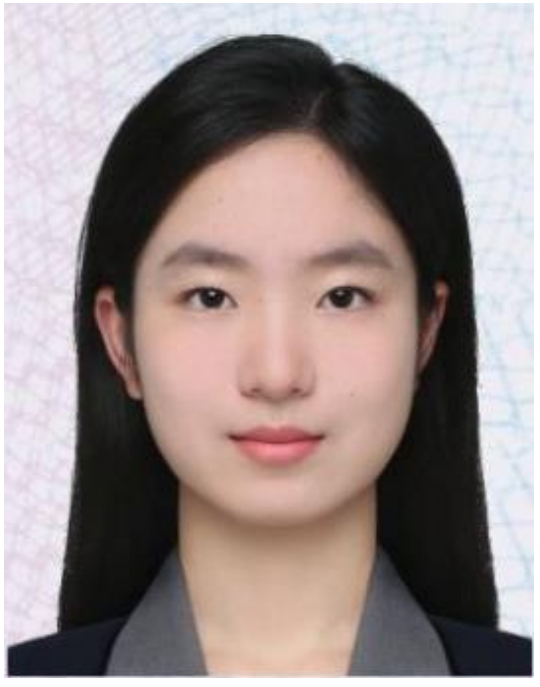

**Yuting FENG** is now studying for a B.E. degree in electronic information engineering at Yangzhou University in China. Her research interests include image processing and machine learning.

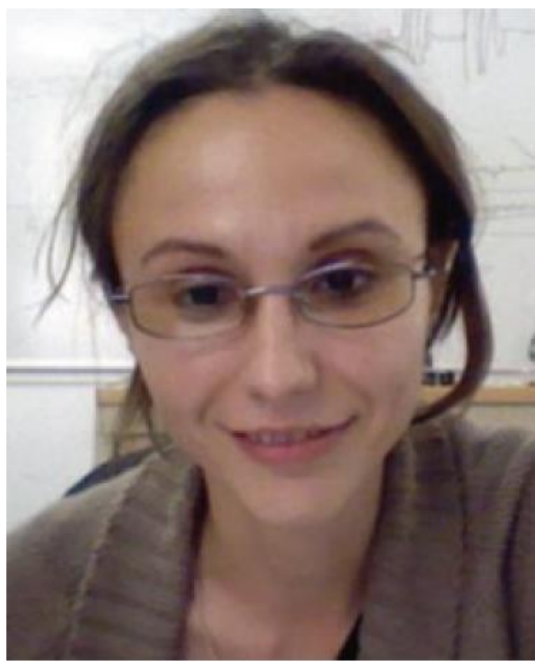

**Maria TROCAN** received a M.E. in Electrical Engineering and Computer Science from the Politehnica University of Bucharest, a Ph.D. in Signal and Image Processing from Telecom ParisTech, and the Habilitation to Lead Researches (HDR) from Pierre & Marie Curie University (Paris 6). She has joined Joost - Netherlands, where she worked as a research engineer involved in the design and development of video transcoding systems. She is firstly Associate Professor, then Professor at Institut Superieur d'Electronique de Paris (ISEP). She is an Associate Editor for the Springer Journal on Signal, Image and Video Processing and a Guest Editor for several journals (Analog Integrated Circuits and Signal Processing, IEEE Communications Magazine, etc.). She is an active member of IEEE France and served as a counselor for the ISEP IEEE Student Branch, IEEE France Vice-President responsible for Student Activities, and IEEE Circuits and Systems Board of Governors member, as Young Professionals representative. Her current research interests focus on image and video analysis & compression, sparse signal representations, machine learning, and fuzzy inference.

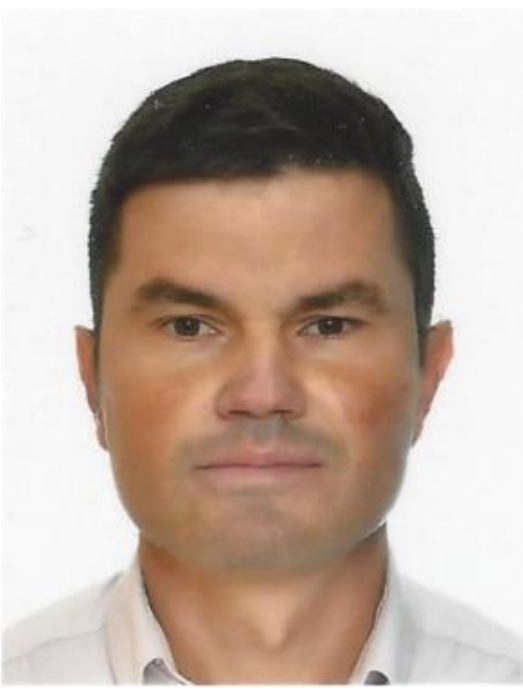

**Dimitri GALAYKO** received a bachelor's degree from Odessa State Polytechnic University in Ukraine, a master's degree from the Institute of Applied Sciences of Lyon in France, and a Ph.D. degree from University of Lille in France. He made his Ph.D. thesis at the Institute of Microelectronics and Nanotechnologies. His Ph.D. dissertation was on the design of micro-electromechanical silicon filters and resonators for radio communications. He is a Professor at the LIP6 research laboratory of Sorbonne University in France. His research interests include the study, modeling, and design of nonlinear integrated circuits for sensor interfaces and mixed-signal applications. His research interests also include machine learning and fuzzy computing.

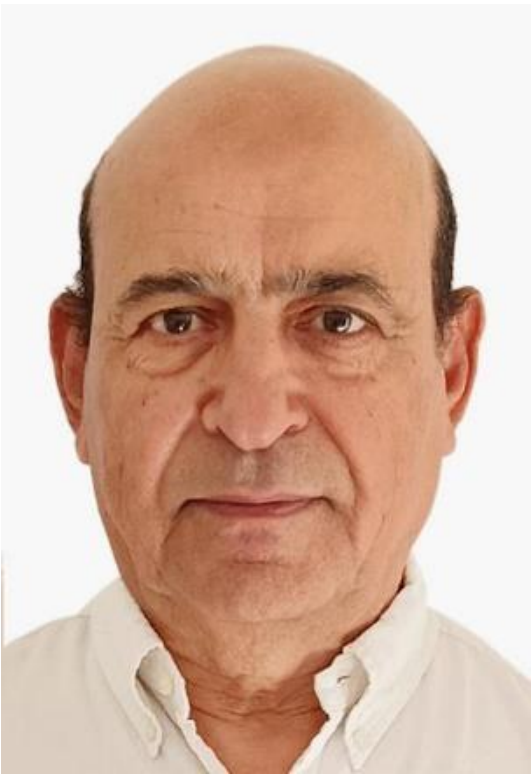

Amara AMARA (Senior Member, IEEE) received the HDR (Confirmation of Leading Research Capabilities) degree from Evry University and the Ph.D. degree from Paris VI University in 1989. In 1988, he joined the IBM Research and Development Laboratory, Corbeil-Essonnes, as a Visiting Researcher, where he was involved in SRAM memory design with advanced CMOS technologies. In 1992, he joined the Paris Institute for Electronics (ISEP) in charge of the Microelectronics Laboratory, where he headed a joint team (Paris VI and ISEP) involved in highspeed GaAs VLSI circuit design. He established the LISTE Laboratory composed of more than 40 researchers in the fields of micro and nano electronics, image and signal processing, and big data processing and analysis. He is the coauthor of three books on Molecular Electronics, Double Gate Devices and Circuits, and Emerging Technologies. He is the author or coauthor of more than 100 conference papers and journal articles. He was also the advisor of more than 20 Ph.D. students. He was the Deputy Managing Director of ISEP in charge of Research and International Cooperation since March 2017. He then joined in June 2017 "Terre des hommes" (Tdh) an international NGO specialized in child protection where he was in charge of ICT for Development (ICT4D) and Artificial Intelligence for Health. He was the President of the CAS Society from 2020 to 2021. He is the former IEEE France Section Chair and the Co-Founder and the Chair of the IEEE France CASS Chapter.

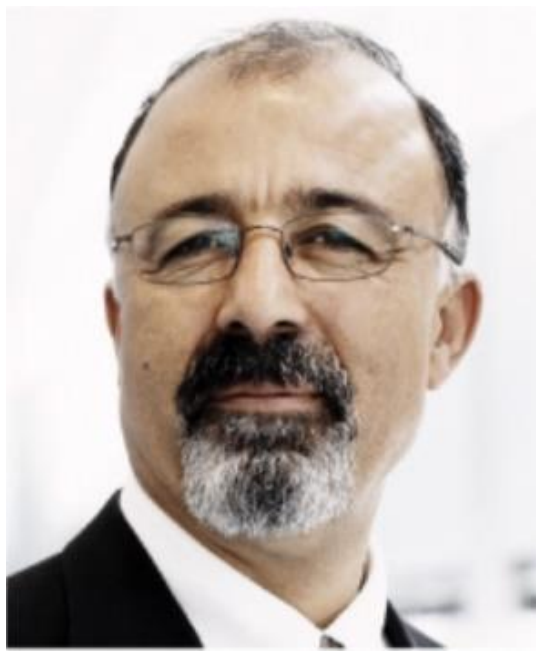

**Mohamad SAWAN** (Fellow, IEEE) received a Ph.D. degree in electrical engineering from the University of Sherbrooke, Sherbrooke, QC, Canada, in 1990. He was a Chair Professor awarded with the Canada Research Chair in Smart Medical Devices (2001–2015) and was leading the Microsystems Strategic Alliance of Quebec - ReSMiQ (1999–2018). He is a Professor of Microelectronics and Biomedical Engineering, in leave of absence from Polytechnique Montréal, Canada. He joined Westlake University, Hangzhou, China, in January 2019, where he is a Chair Professor, Founder, and Director of the Center for Biomedical Research And Innovation (CenBRAIN). He has published more than 800 peer-reviewed articles, two books, ten book chapters, and 12 patents. He founded and chaired the IEEE-Solid State Circuits Society Montreal Chapter (1999–2018) and founded the Polystim Neurotech Laboratory, Polytechnique Montréal (1994–present), including two major research infrastructures intended to build advanced Medical devices. He is the Founder of the International IEEE-NEWCAS Conference, and the Co-Founder of the International IEEE-BioCAS, ICECS, and LSC conferences. He is a Fellow of the Royal Society of Canada, a Fellow of the Canadian Academy of Engineering, and a Fellow of the Engineering Institutes of Canada. He is also the "Officer" of the National Order of Quebec. He has served as a member of the Board of Governors (2014–2018). He is the Vice-President of Publications (2019–present) of the IEEE CAS Society. He received several awards, among them the Queen Elizabeth II Golden Jubilee Medal, the Barbara Turnbull 2003 Award for spinal cord research, the Bombardier and Jacques-Rousseau Awards for academic achievements, the Shanghai International Collaboration Award, and the Medal of Merit from the President of Lebanon for his outstanding contributions. He was the Deputy Editor-in-Chief of the IEEE TRANSACTIONS ON CIRCUITS AND SYSTEMS-II: EXPRESS BRIEFS (2010–2013); the Co-Founder, an Associate

Editor, and the Editor-in-Chief of the IEEE TRANSACTIONS ON BIOMEDICAL CIRCUITS AND SYSTEMS; an Associate Editor of the IEEE TRANSACTIONS ON BIOMEDICALS ENGINEERING; and the International Journal of Circuit Theory and Applications.